\documentclass{emulateapj} 

\usepackage {lscape, graphicx}
\gdef\rfUV{U-V}
\gdef\rfVJ{V-J}
\gdef\1054{MS\,1054--03}

\def\farcs{\hbox{$.\!\!^{\prime\prime}$}}
\def\simgeq{{\raise.0ex\hbox{$\mathchar"013E$}\mkern-14mu\lower1.2ex\hbox{$\mathchar"0218$}}} 

\begin {document}

\title {Color Distributions, Number and Mass Densities of Massive Galaxies at $1.5<z<3$: Comparing Observations with Merger Simulations}

\author{Stijn Wuyts\altaffilmark{1,2}, Marijn Franx\altaffilmark{3}, Thomas J. Cox\altaffilmark{1,2}, Natascha M. F\"{o}rster Schreiber\altaffilmark{4}, Christopher C. Hayward\altaffilmark{1}, Lars Hernquist\altaffilmark{1}, Philip F. Hopkins\altaffilmark{5}, Ivo Labb\'{e}\altaffilmark{6}, Danilo Marchesini\altaffilmark{7}, Brant E. Robertson\altaffilmark{8,9}, Sune Toft\altaffilmark{10,11}, Pieter G. van Dokkum\altaffilmark{7}}
\altaffiltext{1}{Harvard-Smithsonian Center for Astrophysics, 60 Garden Street, Cambridge, MA 02138}
\altaffiltext{2}{W. M. Keck Postdoctoral Fellow}
\altaffiltext{3}{Leiden University, Leiden Observatory, P.O. Box 9513, NL-2300 RA, Leiden, The Netherlands.}
\altaffiltext{4}{MPE, Giessenbackstrasse, D-85748, Garching, Germany}
\altaffiltext{5}{Department of Astronomy, University of California Berkeley, Berkeley, CA 94720}
\altaffiltext{6}{Carnegie Observatories, 813 Santa Barbara Street, Pasadena, CA 91101; Hubble Fellow}
\altaffiltext{7}{Department of Astronomy, Yale University, New Haven CT 06520-8101}
\altaffiltext{8}{Kavli Institute for Cosmological Physics, and Department of Astronomy and Astrophysics, University of Chicago, Chicago, IL 60637}
\altaffiltext{9}{Enrico Fermi Institute, Chicago, IL 60637; Spitzer Fellow}
\altaffiltext{10}{Dark Cosmology Centre, Niels Bohr Institute, University of Copenhagen, Juliane Maries Vei 30, DK-2100 Copenhagen, Denmark}
\altaffiltext{11}{European Southern Observatory, Karl-Schwarzschild-Str. 2, D-85748 Garching bei M\"{u}nchen, Germany}

\begin{abstract}
We present a comparison between the observed color distribution,
number and mass density of massive galaxies at $1.5<z<3$ and a model
by Hopkins et al. that relates the quasar and galaxy population on the
basis of gas-rich mergers.  In order to test the hypothesis that
quiescent red galaxies are formed after a gas-rich merger involving
quasar activity, we confront photometry of massive ($M>4 \times
10^{10}\ M_{\sun}$) galaxies extracted from the FIRES, GOODS-South,
and MUSYC surveys, together spanning an area of 496 arcmin$^2$, with
synthetic photometry from hydrodynamical merger simulations.  As in
the Hopkins et al. (2006b) model, we use the observed quasar
luminosity function to estimate the merger rate.  We find that the
synthetic $U-V$ and $V-J$ colors of galaxies that had a quasar phase
in their past match the colors of observed galaxies that are best
characterized by a quiescent stellar population.  At $z \sim 2.6$, the
observed number and mass density of quiescent red galaxies with $M > 4
\times 10^{10}\ M_{\sun}$ is consistent with the model in which every
quiescent massive galaxy underwent a quasar phase in the past.  At $z
\sim 1.9$, 2.8 times less quiescent galaxies are observed than
predicted by the model as descendants of higher redshift quasars.  The
merger model also predicts a large number and mass density of galaxies
undergoing star formation driven by the merger.  We find that the
predicted number and mass density accounts for 30-50\% of the observed
massive star-forming galaxies.  However, their colors do not match
those of observed star-forming galaxies.  In particular, the colors of
dusty red galaxies (accounting for 30-40\% of the massive galaxy
population) are not reproduced by the simulations.  Several possible
origins of this discrepancy are discussed.  The observational
constraints on the validity of the model are currently limited by
cosmic variance and uncertainties in stellar population synthesis and
radiative transfer.
\end{abstract}


\section {Introduction}
\label{intro.sec}

In recent years, deep near- and mid-infrared observations have
revealed significant populations of red galaxies at redshifts $z \sim
2$ and above (Franx et al. 2003; Daddi et al. 2004; Yan et al. 2004).
The population of Distant Red Galaxies (DRGs), selected by the simple
observed color criterion $J-K>2.3$, makes up 66\% in number and 73\%
in mass of the $2<z<3$ galaxy population at the high mass end
($M>10^{11}\ M_{\sun}$, van Dokkum et al. 2006, see also Marchesini et
al. 2007).  Probing to lower masses, Wuyts et al. (2007) found that
the lower mass galaxies at redshifts $2<z<3.5$ have bluer rest-frame
$U-V$ colors compared to the most massive galaxies.  A substantial
fraction of the massive red galaxies at high redshift are best
characterized by a quiescent stellar population on the basis of their
broad-band SEDs (Labb\'{e} et al. 2005; Wuyts et al. 2007) and the
presence of a Balmer/4000\AA\ break and absence of emission lines in
their rest-frame optical spectra (Kriek et al. 2006).

Any satisfying theory of galaxy formation has to account for the
presence and abundance of these massive red galaxies in the early
universe, a condition that was by no means met by the state-of-the-art
hierarchical galaxy formation models at the time of their discovery
(Somerville 2004).

In the meantime, merger scenarios involving AGN activity have been
invoked by semi-analytic models (Granato et al. 2004; Croton et
al. 2006; Bower et al. 2006; De Lucia \& Blaizot 2007; Somerville et
al. 2008) and hydrodynamical simulations (Springel et al. 2005a; Di
Matteo et al. 2005) to explain simultaneously the mass build-up of
galaxies and the shutdown of star formation.  Such an evolutionary
scenario predicts an obscured (and thus red) star-burst phase and ends
with a quiescent (and thus red) remnant galaxy (e.g., Hopkins et
al. 2006a).  Observational support for the connection between
dust-enshrouded starbursts, merging, and AGN activity from samples of
nearby Ultra-Luminous Infrared Galaxies (ULIRGs) dates from as early
as Sanders et al. (1988).  Furthermore, the observed relation between
the supermassive black hole (SMBH) mass and the mass (Magorrian et
al. 1998) or the velocity dispersion (Ferrarese \& Merritt 2000;
Gebhardt et al. 2000) of their host suggests that black hole and
galaxy growth are intimately connected.  This scaling relation can be
reproduced by merger simulations with implemented AGN feedback (Di
Matteo et al. 2005; Robertson et al. 2006b).

Motivated by the observed and simulated correlations between the
properties of SMBHs and their hosts, Hopkins et al. (2006b) used the
observed quasar luminosity function to derive the galaxy merger rate
as a function of mass.  This paper uses the merger rate function
derived by this model in combination with hydrodynamical SPH
simulations by Robertson et al. (2006a) and T. J. Cox to predict the
color distribution, number and mass density of massive galaxies in the
redshift range $1.5<z<3$ under the assumption that each galaxy once
had or will undergo a quasar phase.  We compare the results to
mass-limited samples in the same redshift interval, extracted from the
multi-wavelength surveys FIRES (Franx et al. 2000; Labb\'{e} et
al. 2003; F\"{o}rster Schreiber et al. 2006a), GOODS-South (Giavalisco
et al. 2004; Wuyts et al. 2008), and MUSYC (Quadri et al. 2007).

The model we analyze in this paper resides in a much larger context
that predicts that morphological transformations, starbursts, quasars,
and the growth of structure are driven by galaxy mergers.  This model
has been well calibrated to many observations at low redshift (e.g.,
Jonsson et al. 2006 and Rocha et al. 2008 on attenuation of local
spiral galaxies and mergers; Hopkins et al. 2008, 2009 on the
structure of local ellipticals) and high redshift (e.g., Younger et
al. 2009 and Narayanan et al. 2009 on the infrared output of
high-redshift (Ultra-)Luminous InfraRed Galaxies and Sub-Millimeter
Galaxies).  Here, we study the epoch of $1.5<z<3$, when AGN and star
formation activity was at its peak, and consider specifically whether
the observations of massive galaxies can be understood within this
context.  The comparison aims to shed light on the nature of massive
galaxies and their evolutionary history, as well as identify where
refinements to the model are needed.

We give an overview of the observations and simulations in
\S\ref{overview_obs.sec} and \S\ref{overview_sim.sec} respectively.
Next, the sample selection is explained in \S\ref{selection.sec}.
\S\ref{methodology.sec} addresses the methodology to populate a model
universe with the binary merger simulations in order to predict number
densities, mass densities, and color distributions.  We compare the
predicted abundance of massive galaxies by the model to the
observations in \S\ref{all_density.sec}.  The optical and
optical-to-NIR color distribution of observed and simulated massive
galaxies will be addressed in \S\ref{analysis_col.sec}, followed by a
discussion of their specific star formation rates (\S\ref{sSFR.sec})
and of the number and mass density of quiescent and star-forming
massive galaxies in \S\ref{analysis_nrho.sec}.  We briefly compare
observed and modeled pair statistics (\S\ref{pairstat.sec}) and
address a few caveats on the observational and modeling results in
\S\ref{comments.sec}.  Finally, we summarize results in
\S\ref{summary.sec}.

We work in the AB magnitude system throughout this paper and adopt a
$H0 = 70\ km\ s^{-1}\ Mpc^{-1}$, $\Omega_M = 0.3$, $\Omega_{\Lambda} =
0.7$ cosmology.

\section {Overview of the observations}
\label{overview_obs.sec}

\subsection {Fields, coverage, and depth}
\label{data.sec}

We combine $K_s$-band selected catalogs of three different surveys:
FIRES, GOODS-South, and MUSYC.  The reduction and photometry of the
FIRES observations of the Hubble Deep Field South (HDFS) is presented
by Labb\'{e} et al. (2003) and was later augmented with IRAC data.
The field reaches a $K_s$-band depth of 25.6 mag (AB, $5\sigma$ for
point sources) and covers 5 arcmin$^2$.  It was exposed in the WFPC2
$U_{300}$, $B_{450}$, $V_{606}$, $I_{814}$ passbands, the ISAAC $J_s$,
$H$, and $K_s$ bands, and the 4 IRAC channels.  Following similar
procedures, a $K_s$-band selected catalog for the FIRES \1054 field
was constructed by F\"{o}rster Schreiber et al. (2006a).  The field,
covering 19 arcmin$^2$, has a $K_s$-band depth of 25 mag (AB,
$5\sigma$ for point sources).  The catalog comprises FORS1 $U$, $B$,
and $V$, WFPC2 $V_{606}$, and $I_{814}$, ISAAC $J$, $H$, and $K_s$,
and IRAC 3.6 $\mu$m - 8.0 $\mu$m photometry.

Over a significantly larger area (114 arcmin$^2$), but to a shallower
depth, a $K_s$-band selected catalog, dubbed FIREWORKS, was
constructed based on the publicly available GOODS-South data (Wuyts et
al. 2008).  The variations in exposure time and observing conditions
between the different ISAAC pointings lead to an inhomogeneous depth
over the whole GOODS-South field.  The 90\% completeness level in the
$K_s$-band mosaic is reached at an AB magnitude of $K_{tot, AB} =
23.7$.  The photometry was performed in an identical way to that of
the FIRES fields, allowing a straightforward combination of the three
fields.  The included passbands are the ACS $B_{435}$, $V_{606}$,
$i_{775}$, and $z_{850}$ bands, the ISAAC $J$, $H$, and $K_s$ bands,
and the 4 IRAC channels.  We also use the ultradeep MIPS 24 $\mu$m (20
$\mu$Jy, 5$\sigma$) imaging of the GOODS-South field.  As for the IRAC
bands, we used the information on position and extent of the sources
from the higher resolution $K_s$-band image to reduce confusion
effects on the 24 $\mu$m photometry (Labb\'{e} et al. in preparation).

Finally, we complement the FIRES and GOODS-South imaging with
optical-to-MIR observations of the MUSYC HDFS1, HDFS2, 1030, and 1255
fields for parts of our analysis.  The $K_s$-band selected catalogs,
augmented with IRAC photometry, are presented by Marchesini et al. (2008).
Together, the MUSYC fields span an area of 358 arcmin$^2$.  They reach
the 90\% completeness level at $K_{tot, AB} = 22.7$.  Given their
shallower depth, they will only be used in the analysis of the most
massive ($M > 10^{11}\ M_{\sun}$) high-redshift galaxies.

\subsection {Redshifts and rest-frame photometry}
\label{z_and_rfphot.sec}

Despite the large number of spectroscopic campaigns in the GOODS-South
and FIRES fields, the fraction of $K_s$-selected $1.5<z<3$ galaxies
that is spectroscopically confirmed is only 9\%.  The fraction drops
to 3\% when the MUSYC fields are included.  Therefore, a reliable
estimate of the photometric redshift is crucial in defining robust
samples of massive high-redshift galaxies.

Wuyts et al. (2008) used the EAZY photometric redshift code by
Brammer, van Dokkum \& Coppi (2008) to fit a non-negative linear
combination of galaxy templates to the FIREWORKS $U_{38}$-to-$8 \mu$m
spectral energy distributions of galaxies in the GOODS-South field.
We applied an identical procedure to galaxies in the FIRES fields.
The template set was constructed from a large number of P\'{E}GASE
models (Fiox \& Rocca-Volmerange 1997).  It consists of 5 principal
component templates that span the colors of galaxies in the
semi-analytic model by De Lucia \& Blaizot (2007), plus an additional
template representing a young (50 Myr) and heavily obscured ($A_V =
2.75$) stellar population to account for the existence of dustier
galaxies than present in the semianalytic model.  A template error
function was applied to downweight the rest-frame UV and rest-frame
NIR during the fitting procedure.  The $K_s$-band magnitude $m_0$ was
used as a prior in constructing the redshift probability distribution
$p(z|C,m_0)$ for a galaxy with colors C.  We adopt the value $z_{\rm
mp}$ of the redshift marginalized over the total probability
distribution,
\begin {equation}
z_{\rm mp} = \frac{\int z\ p(z|C,m_0)\ dz} {\int p(z|C,m_0)\ dz},
\end {equation}
as best estimate of the galaxy's redshift.

The uncertainties in the photometric redshifts were determined from
Monte Carlo simulations.  For each galaxy, a set of 100 mock SEDs was
created by perturbing each flux point according to its formal error
bar, and repeating the $z_{\rm phot}$ computation.  The lower and upper
error on $z_{\rm phot}$ comprise the central 68\% of the Monte Carlo
distribution.

We tested the quality of the photometric redshifts in two ways.  First
we compare them to the available spectroscopic redshifts in the
$1.5<z<3$ interval, resulting in a normalized median absolute
deviation $\sigma_{NMAD} \left(\frac{z_{\rm phot} - z_{\rm spec}}{1 +
z_{\rm spec}} \right) = 0.075$.  The quality measure $\sigma_{NMAD}$
remains the same when the spectroscopic redshifts in the MUSYC fields
are included or excluded.  Second we tested how well we could recover
the redshift from synthetic broad-band photometry of simulated SPH
galaxies placed at redshifts 1.5 to 3.  We found that the considered
template set performed very well ($\sigma_{NMAD}(\Delta z / (1+z)) =
0.027$).  The scatter in the comparison to spectroscopically confirmed
galaxies is larger than that derived from the simulations.  This is
likely due to the fact that the synthetic photometry is based on the
same stellar population synthesis code as the template set used to
recover the redshifts.  Therefore, the second test only studies the
impact of an unknown star formation history, dust and metallicity
distribution on the derived $z_{\rm phot}$.

We computed the rest-frame photometry by interpolating between
observed bands using the best-fit templates as a guide.  Uncertainties
in the rest-frame colors were derived from the same Monte Carlo
simulations mentioned above, and comprise both a contribution from
photometric uncertainties and from $z_{\rm phot}$ uncertainties.  For
a detailed description, we refer the reader to Rudnick et al. (2003).
We used an IDL implementation of the algorithm by Taylor et al. (2009)
dubbed ``InterRest''.

\subsection {Stellar masses}
\label{masses.sec}

The stellar masses of the observed galaxies in FIRES and GOODS-South
were derived by F\"orster Schreiber et al. (in preparation) following
the procedure described by Wuyts et al. (2007).  The stellar masses of
galaxies in MUSYC were derived with the same method by Marchesini et
al. (2008).  Briefly, we fit BC03 templates to the optical-to-8 $\mu$m
SED with the HYPERZ stellar population fitting code, version 1.1
(Bolzonella et al. 2000).  We allow the following star formation
histories: a single stellar population (SSP) without dust, a constant
star formation history (CSF) with dust, and an exponentially declining
star formation history with an {\it e}-folding timescale of 300 Myr
($\tau_{300}$) with dust.  The allowed $A_V$ values ranged from 0 to 4
in step of 0.2, and the attenuation law applied was taken from
Calzetti et al. (2000).  We constrain the time since the onset of star
formation to lie between 50 Myr and the age of the universe at the
respective redshift.  Finally, we scale from a Salpeter (1955) IMF
with lower and upper mass cut-offs of 0.1 M$_{\sun}$ and 100
M$_{\sun}$ to a pseudo-Kroupa IMF by dividing the stellar masses by a
factor of 1.6 (Franx et al. 2008).  Whereas we adopt the BC03 models
for our default analysis, we also performed the SED modeling with
templates from Maraston (2005, hereafter M05) and otherwise identical
settings.  We indicate the results based on M05 models in the plots,
and comment on them where relevant.  On average, estimated stellar
masses are lower by a factor 1.5 when M05 models are used.

\subsection {Star formation rates}
\label{SFR.sec}
We derived estimates of the total (unobscured plus obscured) star
formation rate of the observed galaxies by adding the UV and IR light,
scaled by the calibrations for the local universe (Kennicutt 1998):
\begin {equation}
SFR\ [M_{\sun}\ yr^{-1}] = 1.74 \times 10^{-10} \left( L_{IR} + 3.3 L_{2800}\right) / L_{\sun}
\end {equation}
where the rest-frame luminosity $L_{2800} \equiv \nu L_{\nu}(2800\AA)$
was derived from the observed photometry with the algorithm by Rudnick
et al. (2003).  The total IR luminosity $L_{IR} \equiv L(8-1000$
$\mu$m$)$ was derived from the observed 24 $\mu$m flux density in
combination with the photometric redshift estimate (spectroscopic when
available) following the prescription of Dale \& Helou (2002).  As
best estimate, we adopt the mean of the logarithm of all conversion
factors corresponding to the Dale \& Helou (2002) IR spectral energy
distributions within the range $\alpha=1-2.5$, where $\alpha$
parameterizes the heating intensity level from active ($\alpha=1$) to
quiescent ($\alpha=2.5$) galaxies\footnote[1]{We release the
24$\mu$m-to-SFR conversion in table format on
http://www.strw.leidenuniv.nl/fireworks}.  The variation from
$L_{IR,\alpha=2.5}$ to $L_{IR,\alpha=1}$ is 0.9 dex in the redshift
interval $1.5<z<3$.  Where relevant, we indicate this systematic
uncertainty in the conversion from 24 $\mu$m to $L_{IR}$ and
eventually star formation rate in the plots.

\section {Overview of the simulations}
\label{overview_sim.sec}

We use a set of smoothed particle hydrodynamics (SPH, Lucy 1977;
Gingold \& Monaghan 1977) simulations performed by Robertson et
al. (2006a) and T. J. Cox of co-planar and tilted, equal-mass,
gas-rich ($f_{\rm gas} = 0.8$ at the start of the simulation) mergers
over a range of galaxy masses.  In \S\ref{comments.sec}, the validity
of an equal-mass merger assumption is further discussed in the light
of alternative mechanisms such as minor mergers (\S\ref{minor.sec})
and smooth accretion flows (\S\ref{accretion.sec}).  A description of
the GADGET-2 code used to run the simulations is given by Springel
(2005b).  Springel \& Hernquist (2003) describe the prescriptions for
star formation and supernova feedback.  The interplay between the
supermassive black hole(s) and the environment is discussed by
Springel et al. (2005b).  We refer the reader to Robertson et
al. (2006a) for specifications on this particular set of simulations
and an explanation of how the progenitors were scaled to approximate
the structure of disk galaxies at redshift $z=3$.  The mass resolution
varied from $\log m_i \simeq 5$ per stellar particle for the lowest
mass runs to $\log m_i \simeq 6.5$ per stellar particle for the most
massive mergers.  The photometry of the snapshots, stored with a time
resolution of 14 Myr for the tilted and 70 Myr for the co-planar runs,
was derived in post-processing as described by Wuyts et al. (2009).

Briefly, the total attenuated spectral energy distribution (SED) for a given snapshot consisting of N stellar particles is computed as follows:
\begin {eqnarray}
\displaystyle L_{\rm Att,tot}(\lambda) & = \sum\limits_{i=1}^N & m_i \cdot L_{\rm Int}(age_i, Z_i, \lambda) \nonumber \\
 & & \cdot \exp \left[- N_{\rm H_{i,{\rm los}}} \cdot \frac{Z_{i,{\rm los}}}{Z_{\sun}} \cdot \sigma(\lambda)\right]
\label {phot.eq}
\end {eqnarray}
where m$_i$, age$_i$, and Z$_i$ are, respectively, the mass, age, and
metallicity of stellar particle $i$ that is treated as a single
stellar population.  $L_{\rm Int}$ is the intrinsic (unattenuated) SED
interpolated from a grid of templates from a stellar population
synthesis code.  Here, we use SSP templates from BC03 as default.
Results obtained when using a grid of Maraston (2005, hereafter M05)
SSP templates for different ages and metallicities will be addressed
as well.  For the intrinsic emission $L_{\rm Int}$ of the black hole
particle(s), we scale a luminosity-dependent template SED by the
bolometric black hole luminosity given by the simulation (see Hopkins,
Richards \& Hernquist 2007).  Parameters in Eq.\ \ref{phot.eq} that
are dependent on the line of sight are subscripted with ``los''.  To
each stellar particle, the column density of hydrogen ($N_{\rm H_{i,
{\rm los}}}$) and the average metallicity along the line of sight
($Z_{i, {\rm los}}$) was computed for 100 viewing angles, uniformly
spaced on a sphere.  The optical depth is proportional to this
metallicity-scaled column density, so that $A_B/N_{\rm HI} =
(Z/0.02)(A_B/N_{\rm HI})_{\rm MW}$ where the gas-to-dust ratio of the
Milky Way equals $(A_B/N_{\rm HI})_{\rm MW} = 8.47 \times 10^{-22}\
{\rm cm}^2$.  The wavelength dependence is adopted from an attenuation
law (parameterized by the cross section $\sigma (\lambda)$).  We use
the Calzetti et al. (2000) reddening curve unless mentioned otherwise.
The change in predicted colors when adopting the SMC-like attenuation
law from Pei (1992) will be discussed as well.

\section {Sample selection}
\label{selection.sec}

\begin {figure} [t]
\centering
\plotone{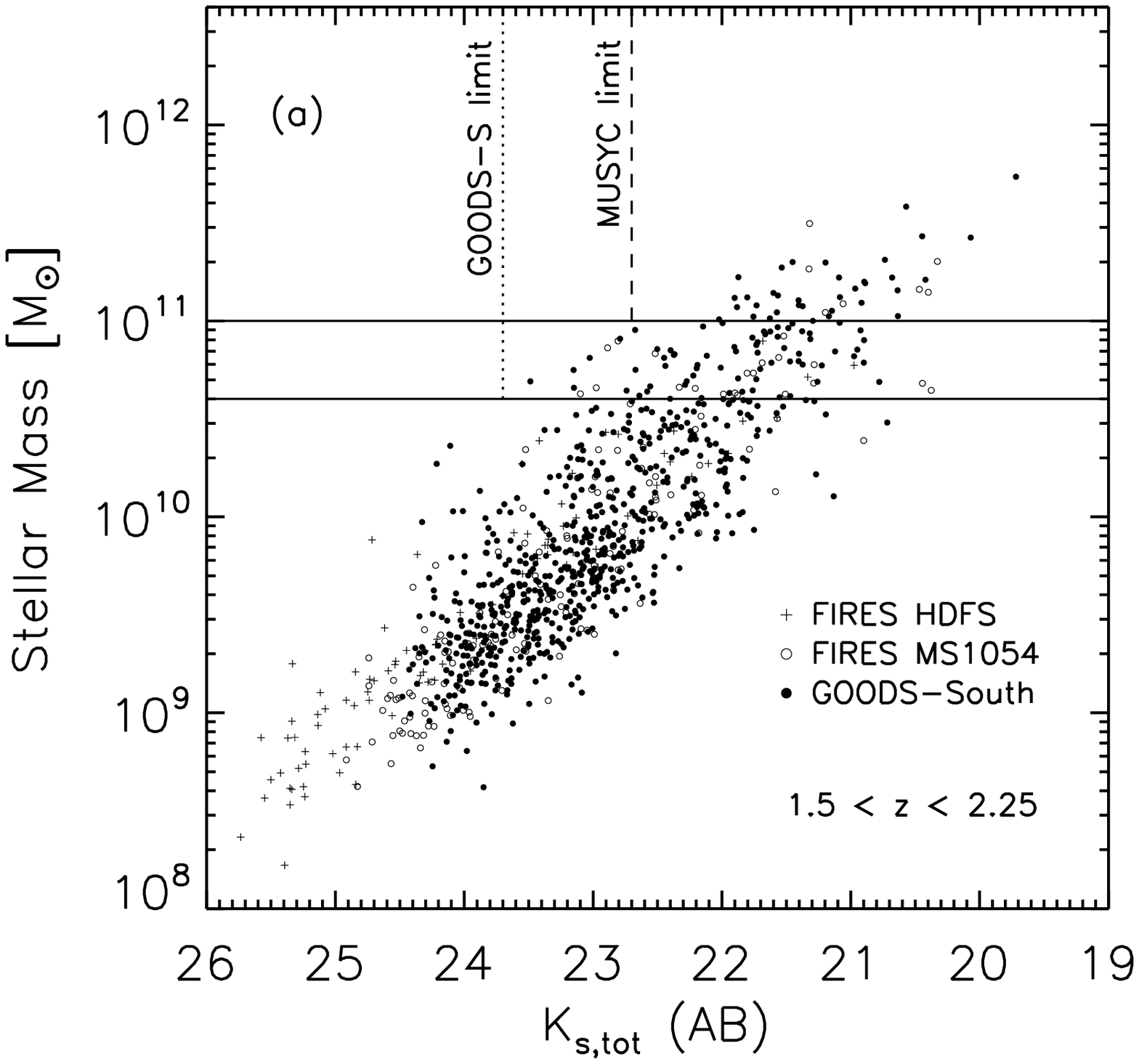} 
\plotone{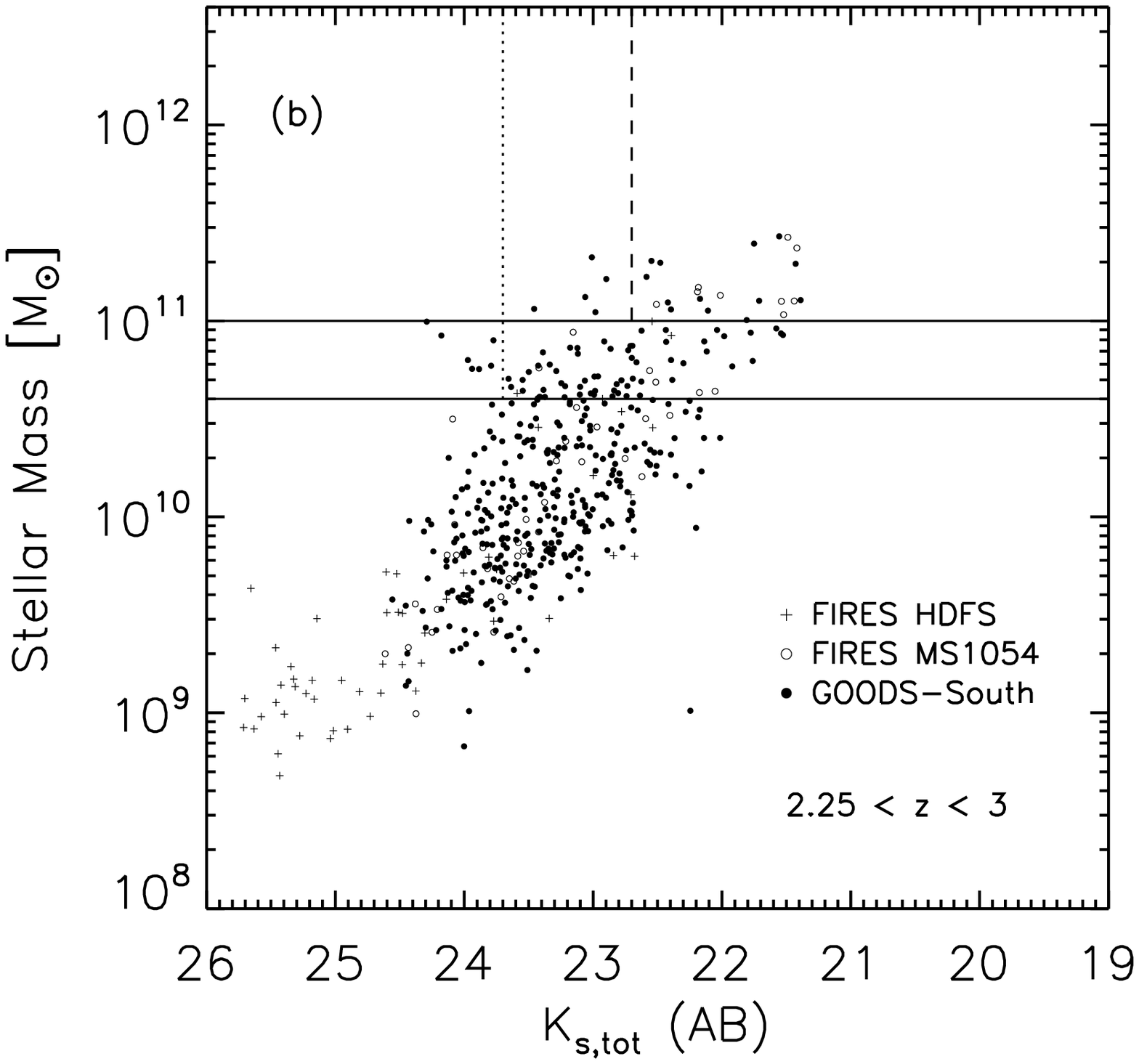} 
\epsscale{1}
\caption{\small The relation between stellar mass and observed total $K_s$
magnitude for galaxies in the FIRES and GOODS-South fields at (a)
$1.5<z<2.25$ and (b) $2.25<z<3$.  The solid lines show the adopted
$\log M > 10.6$ (FIRES+GOODS-South) and $\log M > 11$
(FIRES+GOODS-South+MUSYC) mass limits.  The dotted line indicates the
photometric limit of the GOODS-South imaging.  The dashed line
indicates the approximate limit for the MUSYC fields.  There are few
galaxies with $\log M > 10.6$ and $K_{s, {\rm tot}} > 23.7$, or $\log M > 11$
and $K_{s,{\rm tot}} > 22.7$.  The largest incompleteness correction is
needed for the highest redshift bin in the MUSYC fields.  A fifth of
the $\log M > 11$ galaxies would be undetected by MUSYC, as estimated
from the deeper FIRES+GOODS fields.
\label{sample.fig}}
\end {figure}

Our aim is to compare the color distribution, number and mass density
of mass-limited samples of observed and simulated galaxies.  We choose
the mass-limit such that the observed sample is reasonably complete in
the considered redshift interval, even for the field with the
shallowest $K_s$-band depth from which the sample was drawn.  In order
to optimally exploit the range in area and depth of the considered
surveys, we define two mass-limited samples and divide each in two
redshift bins: $1.5<z<2.25$ and $2.25<z<3$, probing a similar comoving
volume.  The first sample contains galaxies more massive than $\log M
= 10.6$ ($M \simeq 4 \times 10^{10}\ M_{\sun}$) in the FIRES and
GOODS-South fields.  It contains 134 and 106 objects in the low- and
high-redshift bin respectively.  We present the sample in Fig.\
\ref{sample.fig}, where we plot the stellar mass of all FIRES and
FIREWORKS sources that are detected above the $5\sigma$ level in the
respective redshift bin against their total observed $K_s$-band
magnitude.  The stellar mass correlates with the $K_s$-band magnitude,
but a scatter of an order of magnitude is present due to the range in
redshifts and spectral types of the galaxies.  The 90\% completeness
limit ($K_{s,{\rm tot}} = 23.7$) for the GOODS-South field , which is
shallower than the FIRES fields, is indicated with the dotted line.
At $1.5<z<2.25$, no massive ($\log M > 10.6$) galaxies fainter than
$K_{s,{\rm tot}} = 23.7$ are found in the FIRES fields and deeper parts of
the GOODS-South mosaic.  The lowest $K_s$-band signal-to-noise ratio
in the massive galaxy sample is $S/N_{K_s} \simeq 9$, strongly
suggesting that no incompleteness correction is needed to compute the
number and mass density in the $1.5<z<2.25$ redshift bin.  In the
$2.25<z<3$ redshift bin, we find 7 well-detected massive ($\log M >
10.6$) galaxies fainter than the 90\% completeness limit of
GOODS-South.  All of them have $5 < S/N_{K_s} < 10$, whereas the vast
majority (90\%) of massive galaxies with $K_{s,{\rm tot}} < 23.7$ are
detected above the $10\sigma$ level.  Evaluating the fraction of
massive galaxies fainter than $K_{s,{\rm tot}} = 23.7$ in the area that is
sufficiently deep to detect these sources, we estimate the
completeness in the high-redshift bin to be $\sim 93\%$.

In order to reduce the uncertainty from cosmic variance in the derived
number and mass densities, we also compose a sample including the
MUSYC fields, increasing the sampled area by roughly a factor of 3.6.
The shallower depth forces us to restrict the mass limit to $M >
10^{11}\ M_{\sun}$.  The number of objects above this mass limit is
176 at $1.5<z<2.25$ and 71 at $2.25<z<3.0$.  We derive the
completeness in the two redshift intervals using the deeper FIRES and
GOODS-South fields in Fig.\
\ref{sample.fig}.  The dashed line marks the approximate depth (90\%
completeness) for the MUSYC fields.  None of the $1.5<z<2.25$ galaxies
with $\log M > 11$ in the deeper FIRES and GOODS-South fields are
fainter than this limit.  For the $2.25<z<3$ bin, the fraction of
massive galaxies that would be missed by MUSYC increases to 19\%.  In
our analysis, we apply the appropriate incompleteness corrections
unless stated otherwise.  For each considered redshift bin and mass
limit, the sample size decreases by roughly a factor 1.5 when using
M05 models.

\section {Methodology}
\label{methodology.sec}

Large cosmological simulations with sufficient resolution to study the
accretion onto SMBHs are computationally very challenging.  First
attempts were undertaken by Di Matteo et al. (2008), but by and large,
hydrodynamical simulations including a self-consistent treatment of
SMBH growth have only been run with adequate resolution on binary
merger systems (Springel et al. 2005a; Di Matteo et al. 2005;
Robertson et al. 2006a, 2006b; Cox et al. 2006a, 2006b) or as zoom-in
on overdense regions of cosmological N-body simulations at very high
redshift $z \sim 6$ (Li et al. 2007).  In order to confront
observations of $1.5<z<3$ galaxies with the hydrodynamical
simulations, we populate a model universe with the binary mergers
according to a merger rate estimated from the observed quasar
luminosity function following the prescription by Hopkins et
al. (2006b).  As a caveat, we note that this approach does not allow
for a replenishment of the galaxy's gas reservoir by further accretion
from the intergalactic medium (see also \S\ref{minor.sec} and
\S\ref{environment.sec}).  As such, it is not a full cosmological
prediction by itself, but our comparison can be used to see whether
the assumption that the mergers will not experience further infall
leads to reasonable results.

Briefly, the conversion from quasar demographics to galaxy
demographics goes as follows.  From a large set of binary merger
simulations, Hopkins et al. (2006a) determined the distribution of
quasar lifetimes, describing the time
$\frac{dt(L,L_{peak})}{d\log(L)}$ spent by a quasar of peak luminosity
$L_{peak}$ in the luminosity interval $d \log(L)$.  The observed
quasar luminosity function simply corresponds to the convolution of
this differential quasar lifetime with the birth rate
$\dot{n}(L_{peak})$ of quasars with peak luminosity $L_{peak}$:

\begin {equation}
\Phi (L) = \int{\frac{dt(L,L_{peak})}{d\log(L)}\ \dot{n}(L_{peak})\ d\log L_{peak}}
\label {QSO_LF.eq}
\end {equation}

Using a compilation of observed quasar luminosity functions in the
hard X-rays (Ueda et al. 2003), soft X-rays (Hasinger, Miyaji,\&
Schmidt 2005), and optical (Richards et al. 2005), Eq.\
\ref{QSO_LF.eq} was then de-convolved to solve for
$\dot{n}(L_{peak})$.  The relation between peak luminosity of the
quasar and the final black hole mass, derived from the same
simulations, was then adopted to calculate the birth rate of black
holes of a certain final mass $\dot{n}(M_{BH})$.  This function was
then converted to a birth rate of spheroids $\dot{n}(M_{sph})$ as a
function of their final stellar mass using the SMBH-host connection
$M_{BH} = 0.0012\
\frac{(1+z^{2.5})}{\left(1+\left(\frac{z}{1.775}\right)^{2.5}\right)}
\ M_{sph}$ (Hopkins et al. 2007).

\begin {figure} [t]
\centering
\plotone{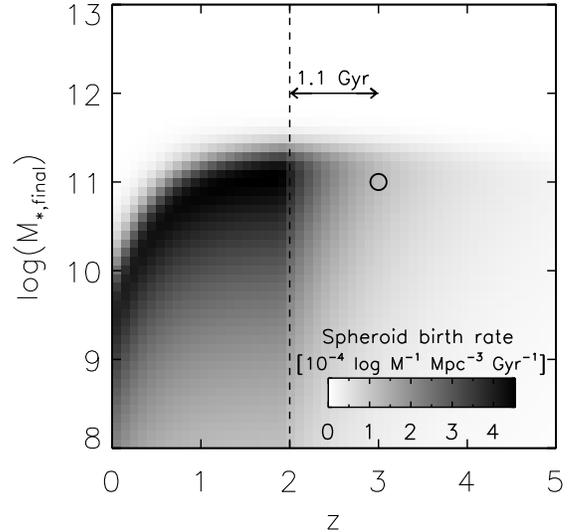} 
\caption{The birth rate of spheroids (in grayscale) as a function of
redshift and final stellar mass as derived from the observed quasar
luminosity function.  The meaning of the time scale arrow and the open
circle is described in the text.  The model by Hopkins et al. (2006b)
assumes that this birth rate equals the merger rate of galaxies.  The
birth rate (i.e., merger rate) reaches a maximum of $4.5 \times
10^{-4}\ \log M^{-1}\ {\rm Mpc}^{-3}\ {\rm Gyr}^{-1}$ at $z \sim 2$.  As time
evolves, the peak of the merger rate function shifts toward lower
mass galaxies.
\label{trigger.fig}}
\end {figure}
The model by Hopkins et al. (2006b) assumes that the birth rate of
spheroids equals the major merger rate of galaxies.  The resulting
merger rate as a function of redshift and final stellar mass is
displayed with grayscales in Fig.\ \ref{trigger.fig} (darker meaning a
higher merger rate).  Its redshift-dependence was derived by
considering observed quasar luminosity functions at a range of
redshifts.  The peak of the merger rate at $z \sim 2$ has a value of
$4.5 \times 10^{-4}\ \log M^{-1}\ Mpc^{-3}\ Gyr^{-1}$.  A clear trend
is visible of mergers occurring in increasingly lower mass systems
as we proceed in time (i.e., to lower redshifts) after this peak.  If
mergers are responsible for a significant part of the growth in
stellar mass, this trend explains at least qualitatively the observed
downsizing of star formation over cosmic time (Cowie et al. 1996).

To evaluate the post-merger (i.e., post-quasar, since the merging event
triggers quasar activity in the simulations) galaxy population at $z
\sim 2$, we integrate the merger rate function from $z=\infty$ to 2
and over the whole stellar mass range.  For example, when the
integration reaches ($M_{*,final} = 10^{11}\ M_{\sun};\ z=3$), marked
by the circle in Fig.\ \ref{trigger.fig}, we compute the photometry of
a merger simulation with a final stellar mass of $10^{11}\ M_{\sun}$
at 1.1 Gyr after the peak of quasar luminosity (the time elapsed
between $z=3$ and $z=2$).  As explained in \S\ref{overview_sim.sec},
we compute the synthetic photometry along 100 lines-of-sight,
uniformly spaced on a sphere.  The number density of galaxies at $z=2$
with colors corresponding to the 100 lines-of-sight is then scaled
according to the value of the merger rate function at ($M_{*,final} =
10^{11}\ M_{\sun};\ z=3$).  Finally, a mass cut is applied to
guarantee an identical selection of observed and simulated galaxies
(\S\ref{biasSED.sec} addresses this step in more detail).

\begin {figure} [t]
\centering
\plotone{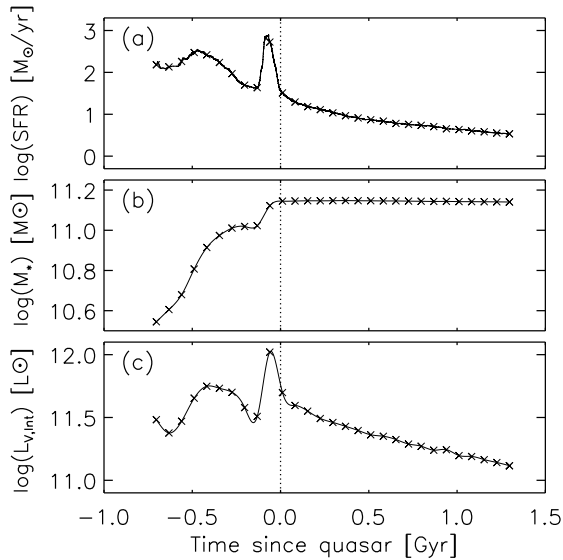} 
\caption{Typical evolution of a merger simulation: (a) star formation
history, (b) history of the mass build-up, and (c) evolution of the
rest-frame $V$-band luminosity.  The dotted line indicates when the peak
in quasar luminosity is reached.  For a detailed description of the
time evolution in these and other parameters (e.g., accretion rate,
quasar luminosity, extinction) we refer the reader to Hopkins et
al. (2006a).
\label{evol.fig}}
\end {figure}
In order to predict the abundance and properties of galaxies at $z
\sim 2$ that have yet to reach their peak in quasar luminosity or did
not even start merging at the evaluated epoch, one can in principle
integrate the merger rate function down to lower and lower redshifts.
How far one integrates beyond the evaluated redshift is a rather
arbitrary choice.  We caution that counting galaxies long before they
will contribute to the quasar luminosity function will lead to large
uncertainties given their unconstrained pre-merger history.  The
typical evolution of a merger simulation is illustrated in Fig.\
\ref{evol.fig} where we plot the star formation rate, stellar mass,
and rest-frame $V$-band luminosity as a function of time since the
peak in quasar luminosity.  We decide to integrate 700 Myr beyond the
evaluated redshift, thus counting both the galaxies that are
undergoing a merger-induced nuclear starburst (sometime between 0 and
200 Myr before the quasar phase) and those with star formation
triggered by the first passage (sometime between 200 and 700 Myr
before the quasar phase).  Hereafter, we will refer to all galaxies in
an evolutionary stage between 0 and 700 Myr before the quasar phase as
merging galaxies.  Such a prediction only counts those galaxies that
will later merge and produce a quasar.  Apart from predicting the
abundance and properties of the post-quasar population, we will thus
be able to constrain how many of the massive star-forming galaxies can
be accounted for by merger-induced star formation.

In the early stages of the merging process, the two progenitor
galaxies may be resolved as two separate objects in the observations.
We simulate this in our model by calculating the projected angular
distance between the central black holes for each snapshot and viewing
angle at the considered redshift.  If the projected separation is
larger than $1\farcs 5$, the 2 progenitors are counted separately,
each containing half the mass.  For smaller projected separations, or
when the black hole particles have merged, we consider the system
unresolved and use the integrated properties in our model predictions.
In cases where the two progenitor components are counted separately,
the merger contributes twice to the number density, but with half the
mass, and may therefore drop out of the mass-limited sample.

Provided the assumption of a one-to-one correspondence between quasars
and major mergers is valid, the formal uncertainty in the merger rate
function presented in Fig.\ \ref{trigger.fig} originates mostly from
the poorly constrained faint end of the observed quasar luminosity
function, where one can assume a pure luminosity evolution or also a
slope evolution.  At the bright end, and therefore for our massive
galaxy samples, the predictions are robust.

\section {The number density, mass density and mass function of galaxies with $\log M > 10.6$ at $1.5<z<3$}
\label{all_density.sec}

\begin {figure} [t]
\centering
\plotone{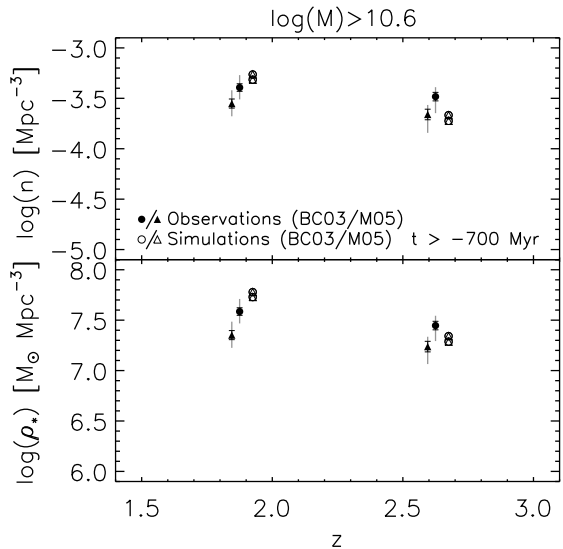} 
\caption{The number and mass density of observed ({\it filled
symbols}; FIRES + GOODS-S) and modeled ({\it empty symbols}) galaxies
with $\log M > 10.6$ as a function of redshift.  The black error bar
represents the Poisson shot noise solely.  The gray error bar accounts
for uncertainties in redshift and mass, and a (dominating)
contribution from cosmic variance.  We find that both the predicted
number and mass densities agree within the error bars with the
observed values.
\label{all.fig}}
\end {figure}

Before analyzing the observed and modeled massive galaxy sample as a
function of color and galaxy type, we consider the overall abundance
of galaxies above $\log M > 10.6$.  We computed the model number and
mass density by integrating the merger rate function to 700 Myr beyond
the evaluated redshift, i.e., including galaxies up to 700 Myr before
the quasar phase.  In Figure\ \ref{all.fig}, the number and mass
densities of galaxies with $\log M > 10.6$ predicted by the model
({\it empty symbols}) are compared against the abundance of observed
galaxies ({\it filled symbols}) above the same mass limit.  Circles
indicate results using our default (BC03) SED modeling.  Triangles
represent the abundances derived using M05 models.  The results are
listed in Table\
\ref{nrho.tab}.  The spread of the empty symbols indicates the freedom
allowed by the model due to the poorly constrained faint end of the
quasar luminosity function.

We considered three sources of error in the observations: Poisson shot
noise, cosmic variance and selection uncertainties stemming from
uncertainties in the redshift and mass estimates of individual galaxies.
The black error bars in Fig.\ \ref{all.fig} indicate the contribution
from Poisson noise, ranging from 8 to 10\%.  We are more severely
limited by cosmic variance.  We follow the method outlined by
Somerville et al. (2004) to calculate the cosmic variance as predicted
from cold dark matter theory for a population with unknown clustering
as a function of its number density and the probed comoving volume of
the sample.  The resulting contribution to the error budget is 29\%
for the $1.5<z<2.25$ and 30\% for the $2.25<z<3$ redshift bin.
Finally, the uncertainties in the individual redshift and mass
determinations propagate into the number and mass density of massive
high-redshift galaxies.  We estimate the contribution to the total
error budget from Monte Carlo simulations.  We constructed 100 mock
catalogs for the FIRES and GOODS-South fields by perturbing the fluxes
so that 68\% of the perturbed values lie within the 1$\sigma$ errors.
We then repeated the computation of photometric redshifts and other
derived properties such as stellar mass.

After constructing the 100 mock catalogs, we apply the same sample
selection (redshift interval, $\log M > 10.6$) and compute the number
and mass density for each of them.  The lower and upper limits
comprising 68\% of the distribution of mock number and mass densities
were added in quadrature to the uncertainty from Poisson shot noise
and cosmic variance, shown with the gray error bar in Fig.\
\ref{all.fig}.  The uncertainty in the number density propagating from
redshift and mass uncertainties for individual objects amounts to 5\%
and 10\% for the low and high-redshift bin.  The contribution to the
uncertainty in the mass density is 6\% and 14\% for the low- and
high-redshift bin respectively.  We conclude that, even with the 138
arcmin$^2$ area of our combined deep fields, cosmic variance is still
the limiting factor for the determination of the number and mass
density of massive high-redshift galaxies.  Furthermore, we note that
the number and mass densities systematically drop by a factor 1.5 and
1.7 respectively when using M05 models.

\begin {figure} [t]
\centering
\plotone{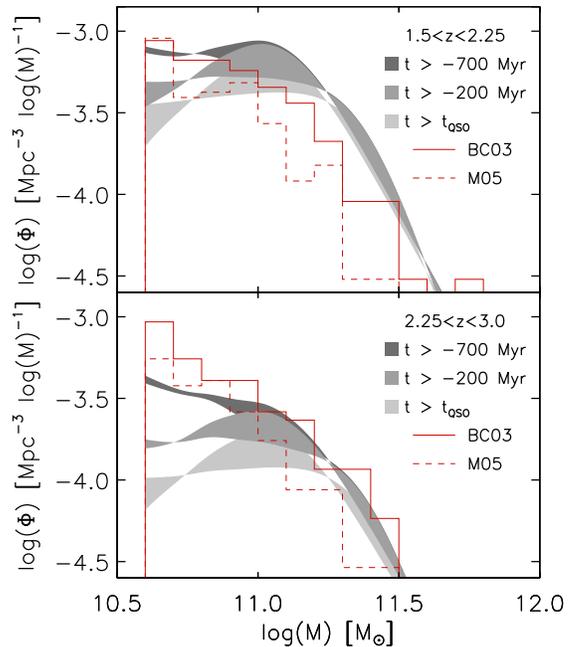} 
\caption{The mass function of observed ({\it red histogram}; FIRES +
GOODS-S) and modeled ({\it gray polygons}) galaxies with $\log M >
10.6$ at redshift $1.5<z<2.25$ ({\it top panel}) and $2.25<z<3.0$
({\it bottom panel}).  Merger remnants alone ({\it $t > t_{QSO}$})
cannot account for the total population of observed galaxies above the
same mass limit.  A better consistency with the observations is
reached when integrating the merger rate function to include galaxies
up to 700 Myr before the quasar phase.
\label{massfunction.fig}}
\end {figure}
Given these uncertainties, Figure\ \ref{all.fig} shows a good
agreement between the model number and mass density for the population
of massive ($\log M > 10.6$) galaxies as a whole and the observations.
Plotting the mass function for the observations ({\it red
histogram}) and the model ({\it dark-gray polygon}) in Figure\
\ref{massfunction.fig}, we find that the comparable abundance of
observed and modeled galaxies still holds when studied as a function
of galaxy mass.  With lighter gray polygons, we illustrate the model
prediction when including only galaxies up to 200 Myr before the
merger ($t >$ -200 Myr) or only merger remnants ($t > t_{QSO}$).  The
width of the polygons reflects the uncertainty in the merger rate
function.  We conclude that merger remnants alone cannot account for
the entire observed massive galaxy population at $1.5<z<3$.  However,
including galaxies with merger-triggered star formation, the mass
function predicted by the model is in good agreement with the
observations.  This encourages a more detailed investigation of the
properties of observed and simulated massive galaxies.

A detailed study of the stellar mass function of observed galaxies
from $z=4.0$ to $z=1.3$, as well as a comprehensive analysis of random
and systematic uncertainties, is presented by Marchesini et
al. (2008).  We note that our abundance estimates of observed massive
galaxies are consistent with those derived from the work of these
authors.  Also, our cosmic variance estimates are consistent with
those empirically derived by Marchesini et al. (2008), who find that
cosmic variance is the dominant source of random uncertainties at
$z<2.5$.

\section {The color distribution of galaxies with $\log M > 10.6$ at $1.5<z<3$}
\label{analysis_col.sec}

\subsection {The $U-V$ color distribution}
\label{UV.sec}
\begin {figure*} [t]
\centering
\epsscale{0.48}
\plotone{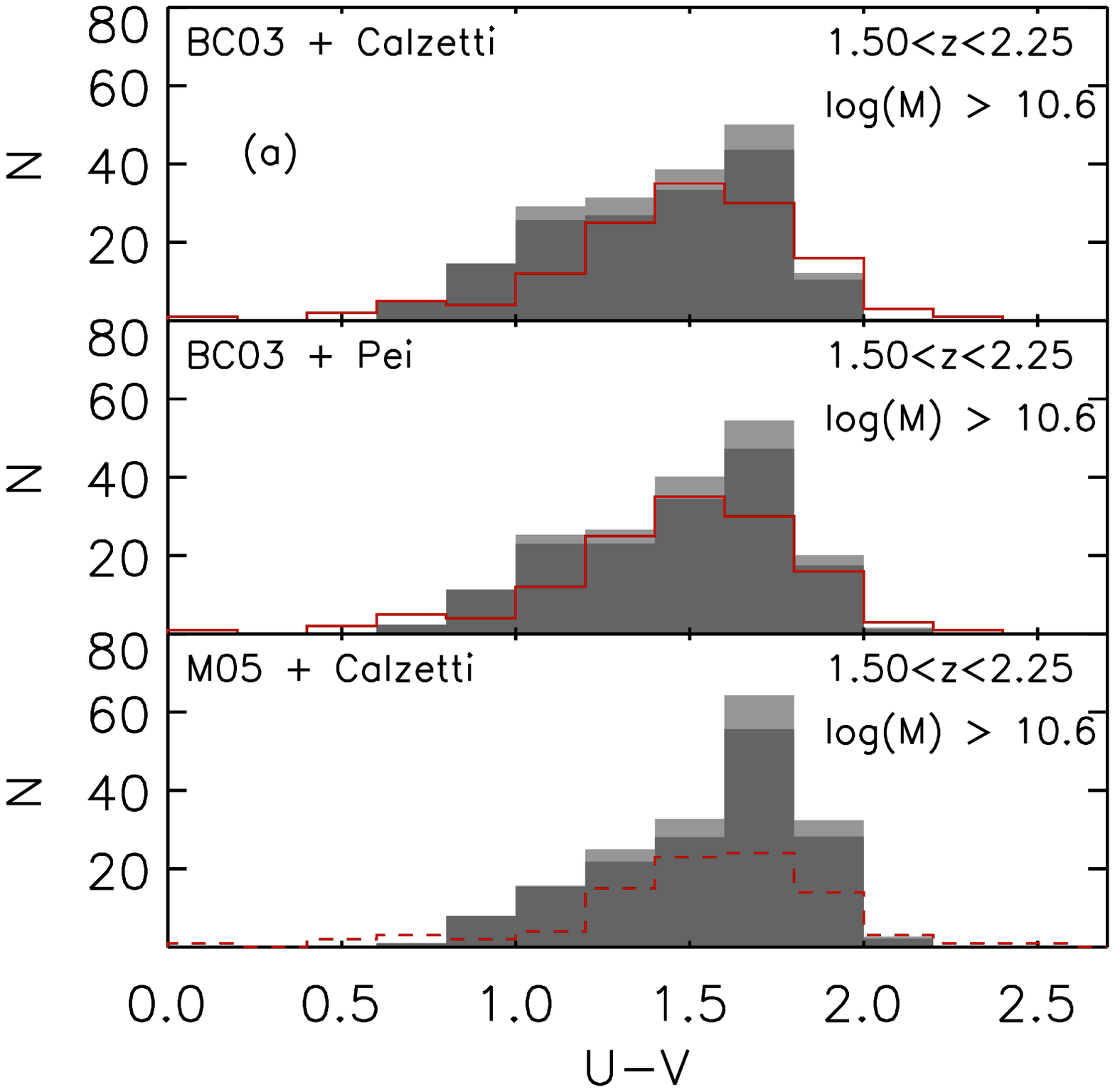}
\plotone{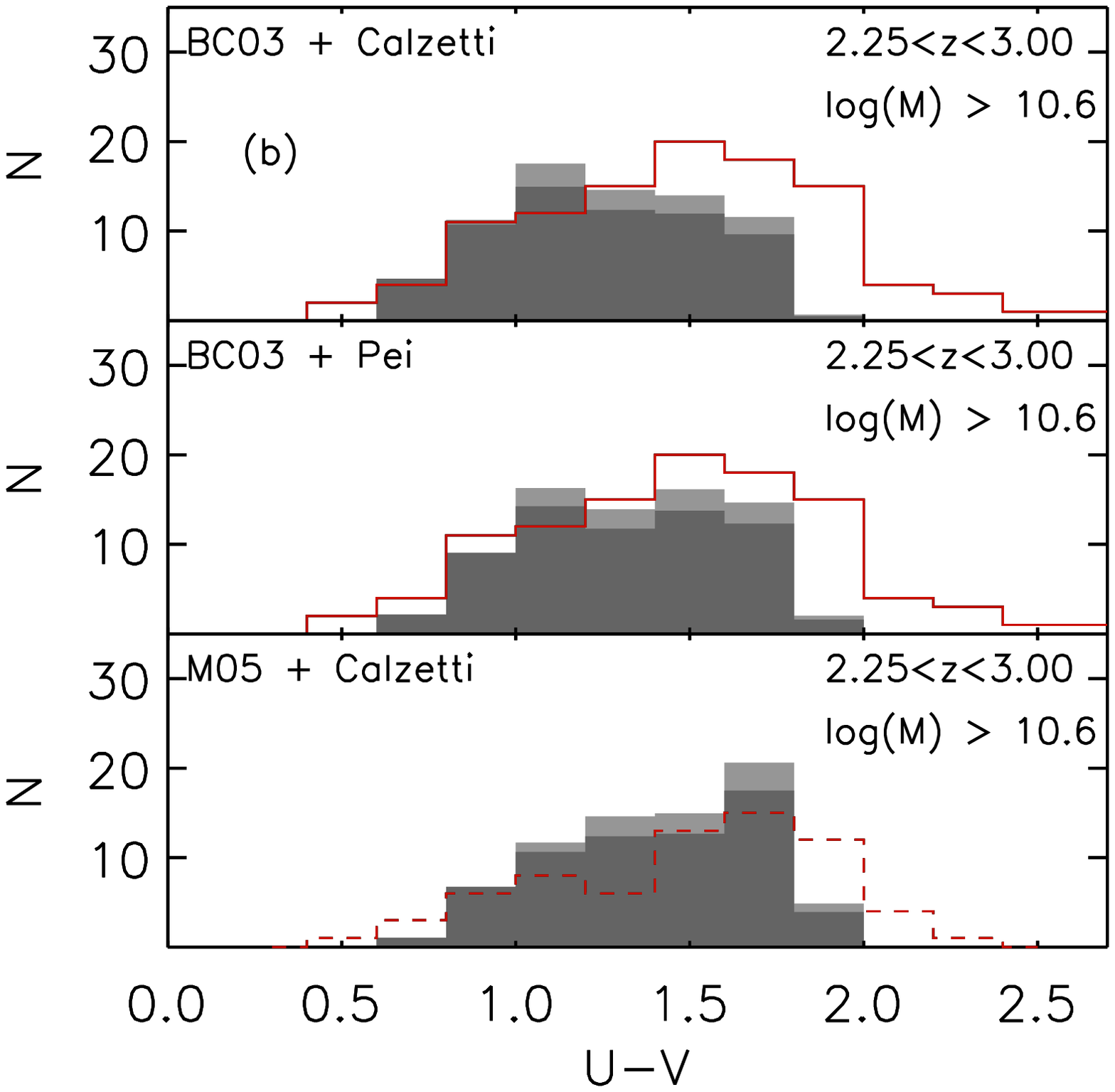} 
\epsscale{1}
\caption{The rest-frame $U-V$ color distribution of observed galaxies
with masses above $\log M = 10.6$ in the FIRES and GOODS-South fields
({\it solid and dashed line for masses based on BC03 and M05
respectively}) for the redshift intervals (a) $1.5<z<2.25$ and (b)
$2.25<z<3$.  Filled histograms are the predicted $U-V$ color
distribution of merging and post-quasar galaxies, scaled to the same
solid angle as the observations.  The light gray top of the model
histogram reflects the uncertainty in the merger rate function.  For a
given redshift interval, the model predictions in the three panels
give an indication of the uncertainty in the synthetic photometry
induced by the choice of attenuation law (Calzetti et al. 2000 versus
the SMC curve from Pei 1992) and the choice of stellar population
synthesis code (BC03 versus M05).  Overall, the predicted color
distribution coincides with that of the observed massive galaxy
sample, with roughly equal numbers.  The red tail of the observed
color distribution at $2.25<z<3$ is not reproduced by the modeled
merger and post-quasar population.
\label{UVhist.fig}}
\end {figure*}

First, we consider the optical color distribution of our sample of
FIRES and FIREWORKS galaxies with $M > 4 \times 10^{10}\ M_{\sun}$.  A
histogram of their rest-frame $U-V$ colors is plotted in red in Fig.\
\ref{UVhist.fig}(a) and Fig.\ \ref{UVhist.fig}(b) for the low- and
high-redshift bin respectively.  No corrections for incompleteness
were applied here, but we remind the reader that those are negligible
for the low-redshift bin and 7\% only for the high-redshift bin.

Massive high-redshift galaxies span a broad $U-V$ color range.  In
both redshift bins, the median color is $U-V=1.5$ and 68\% of the
galaxies lie within the $1.1<U-V<1.8$ interval.

It is interesting to consider whether the simulated descendants and
progenitors of quasars (or rather quasar hosts) above the same mass
limit show colors that are similar and come in numbers comparable to
those of the observed massive galaxy sample.  In this section, we
focus mainly on the first question, but note in passing that we show
the predicted color distribution scaled to the same solid angle as
probed by the FIRES and FIREWORKS observations.  The filled gray
histograms show the synthetic photometry of merger simulations in
either their post-quasar phase or in a phase of at most 700 Myr before
their peak in quasar luminosity.  The numbers at each color are
derived from the observed quasar luminosity function by integrating
the merger rate function from $z=\infty$ to 700 Myr beyond the
evaluated redshift as described in \S\ref{methodology.sec}.  The
colors of different evolutionary phases will be discussed separately
in due time.  The difference between the dark and light gray histogram
reflects the uncertainty in the merger rate function, itself due to
uncertainties in the observed quasar luminosity function.  Apart from
uncertainties in the merger rate function, uncertainties in the
synthetic photometry for a given simulation snapshot contribute to the
total error budget of the model predictions.  To translate the
simulated properties such as age, mass, and metallicity of the stellar
particles to observables, we make use of a stellar population
synthesis code to compute the intrinsic colors and assume an
attenuation law to calculate the dimming and reddening by dust.  We
investigate the dependence on attenuation law empirically by computing
the synthetic photometry using a Calzetti et al. (2000) reddening
curve and the SMC-like reddening curve from Pei (1992).  We note that
the synthetic colors derived with the Milky Way-like attenuation curve
by Pei (1992) lie in between those produced by the two reddening
curves considered here.  This is demonstrated by Wuyts et al. (2009).
Similarly, we test the dependence on adopted stellar population
synthesis templates empirically by computing the synthetic photometry
based on a grid of BC03 single stellar populations (SSPs) and based on
a grid of SSPs by M05.

We note that the choice of attenuation law has a minor effect only on
the $U-V$ color.  The use of M05 templates gives the simulated
galaxies a slightly redder color.  Overall, the same conclusion can be
drawn independent of the choice of attenuation law or stellar
population synthesis code.  Namely, the simulated galaxies with $\log
M > 10.6$ span a color range that reaches from the bluest observed
$U-V$ colors to $U-V \sim 2$.  At $1.5<z<2.25$, the color distribution
resembles remarkably well that of the bulk of the observed massive
galaxies, both in shape and numbers.  We apply a Kolmogorov-Smirnov
(K-S) test, and find that the observed and model color distributions
do not differ at the 5\% significance level.  At $2.25<z<3$, the
predicted model colors do not reach the reddest $U-V$ colors of
observed galaxies above the same mass limit.  A K-S test indicates a
formal difference between the observed and model color distribution.
We do note, however, that the observed sources with $U-V > 2$ have
fairly large uncertainties in their rest-frame optical color
measurement, and are nearly all consistent within 1 $\sigma$ with an
actual color of $U-V < 2$.  The good overall correspondence between
the observed and modeled optical color distributions gives a first
indication that the number of massive post-quasar galaxies plus the
number of galaxies in the process of merging at $1.5<z<3$ as expected
from the observed quasar luminosity function may account for a large
fraction of the observed massive galaxy population in that redshift
range.

\subsection {The $V-J$ color distribution}
\label{VJ.sec}

\begin {figure*} [t]
\centering
\epsscale{0.48}
\plotone{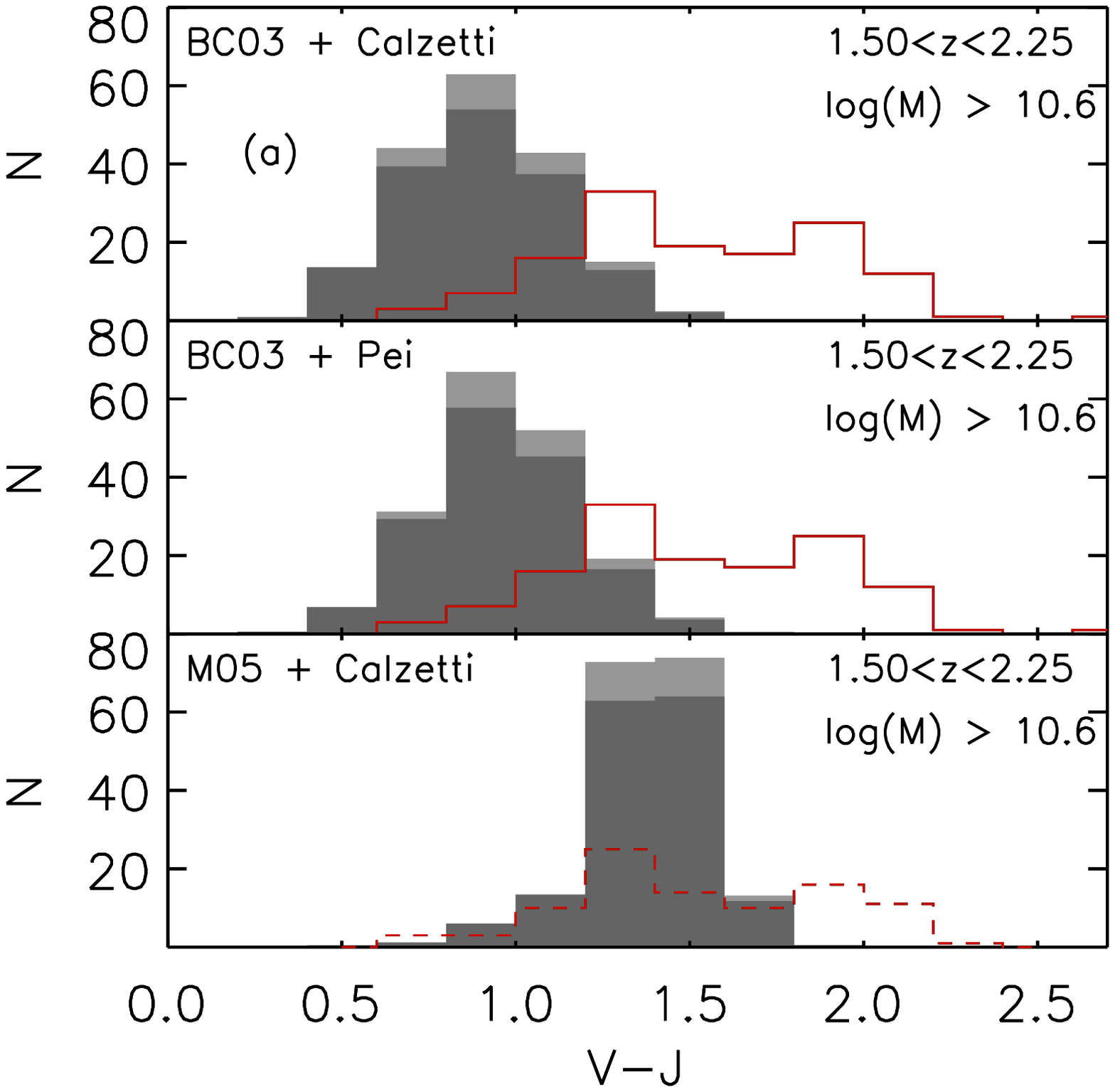}
\plotone{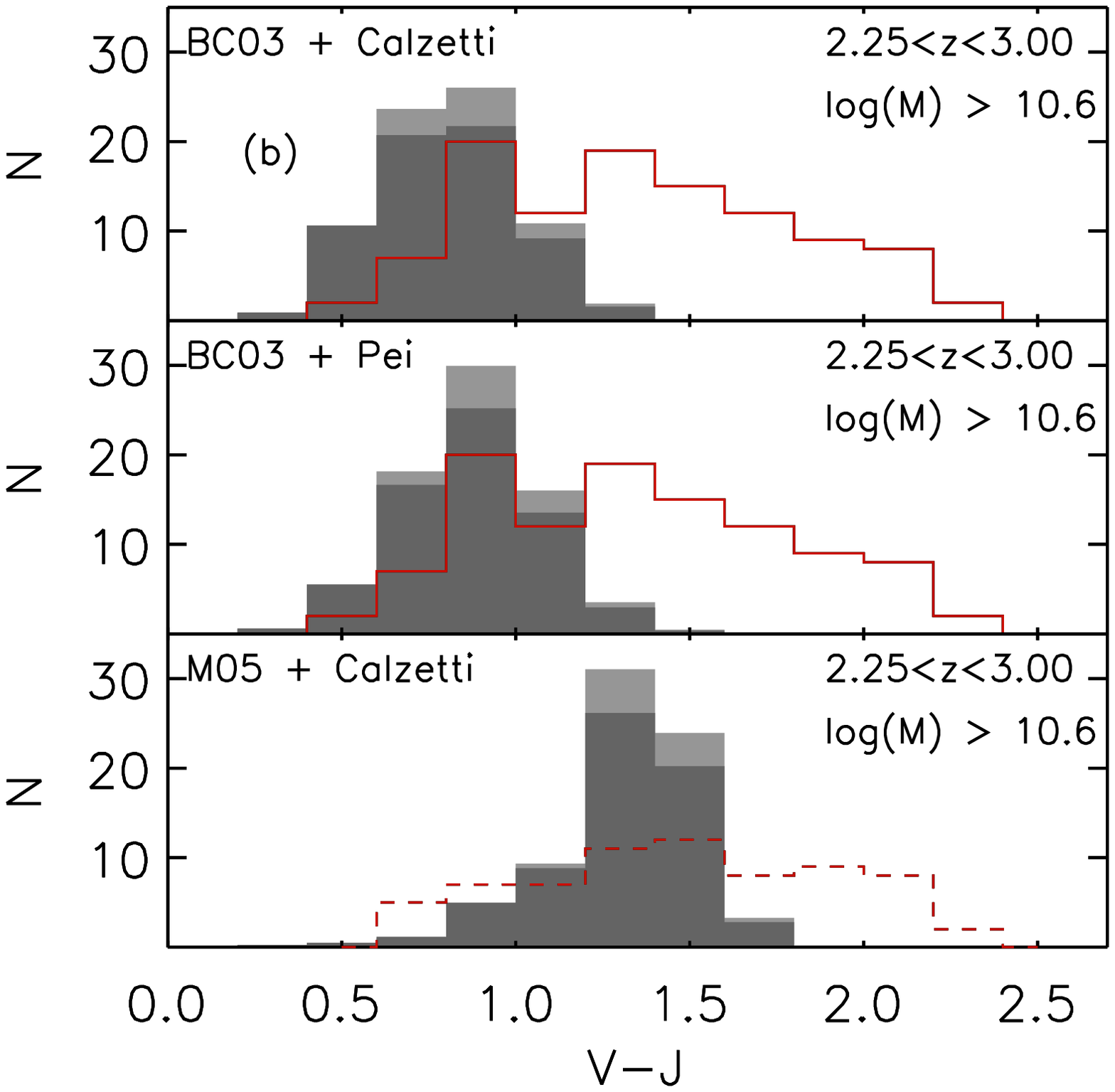} 
\epsscale{1}
\caption{The rest-frame $V-J$ color distribution of observed and simulated 
galaxies with masses above $\log M = 10.6$ for the redshift intervals
(a) $1.5<z<2.25$ and (b) $2.25<z<3$.  Style as in Figure\
\ref{UVhist.fig}.  The model $V-J$ color distribution is only poorly
constrained due to the uncertainties at NIR wavelengths in the stellar
population synthesis codes.  Nevertheless, we can conclude that there
exist massive galaxies with redder $V-J$ colors than those of modeled
merging and post-quasar galaxies.
\label{VJhist.fig}}
\end {figure*}

Turning to longer wavelengths, we now compare the $V-J$ colors predicted
for mergers and merger remnants (i.e., post-quasars) with masses above $\log M =
10.6$ to the color distribution of observed galaxies in the same
redshift interval and above the same mass limit (Fig.\ \ref{VJhist.fig}).

Again, the color distribution of our observed massive galaxy sample
has a large range of colors, reaching from $V-J=0.5$ to $V-J=2.5$.  We
measure a median color of $V-J=1.5$ and 1.3 for the low- and
high-redshift bin respectively.  The central 68\% intervals are
$1.1<V-J<1.9$ and $0.9<V-J<1.8$

As for the $U-V$ color distribution, we find that the adopted
attenuation law has only a minor influence on the model color
distribution, reaching at most shifts of 0.2 mag toward redder $V-J$
colors when the SMC-like reddening curve from Pei (1992) is used
instead of the Calzetti et al. (2000) attenuation law.  Comparing the
model $V-J$ color distribution derived from BC03 or M05 templates
immediately shows that the predictive power of the merger model is
strongly hampered by the uncertainties in the rest-frame NIR
wavelength regime that today's stellar population synthesis codes are
facing.  In the low- and high-redshift bin, the median $V-J$ color of
the model distribution is 0.4 and 0.5 mag redder when using M05 than
when using BC03.  One of the main differences between the BC03 and M05
templates is the treatment of thermally pulsating AGB stars.  M05 uses
the fuel consumption approach instead of the isochrone synthesis
approach that BC03 follow.  In addition, the two models construct the
synthetic populations using different stellar isochrones.  The
combined effect is that M05 finds significantly larger NIR
luminosities for SSPs at ages between 0.2 and 2 Gyr than BC03.  For an
in-depth discussion of the differences between the two codes, we refer
the reader to Maraston (2005) and Maraston et al. (2006).  It is worth
stressing that, irrespective of whether the BC03 or M05 stellar
population synthesis code is used, the red ($V-J > 1.8$) tail of the
observed distribution has no counterparts in the modeled color
distribution of merging and post-merger galaxies.  Conversely, an
excess of galaxies is found at blue ($V-J \sim 0.9$) or intermediate
($V-J \sim 1.4$) optical-to-NIR colors for the BC03 and M05 model
color distributions respectively.  A K-S test shows that the
difference between the distributions is significant at the 99.99\%
level.  Uncertainties in the derived rest-frame colors of observed
galaxies are unlikely to be responsible for the offset.

\subsection{$U-V$ versus $V-J$ color-color distribution}
\label{UVvsVJ.sec}

\subsubsection {Quiescent red galaxies}
\label{UVvsVJdead.sec}

Recently, a diagnostic color-color diagram of observer-frame $I-K$
versus $K$-[4.5 $\mu$m] has been proposed by Labb\'{e} et al. (2005)
to distinguish three basic types of $z>2$ galaxies.  The rest-frame
equivalent of this diagram, $U-V$ versus $V-J$ (hereafter UVJ) allows
a comparison of galaxies over a wider redshift range (Wuyts et
al. 2007; Williams et al. 2009; Labb\'{e} et al. in prep).  First, there
are galaxies with relatively unobscured star formation, such as the
majority of Lyman break galaxies (Steidel et al. 2003) and their lower
redshift BX/BM analogs (Adelberger et al. 2004).  They typically have
young ages and low reddening values, resulting in blue colors, both in
the rest-frame optical and in the rest-frame optical-to-NIR.  Second,
there is a population of star-forming galaxies with much redder
colors, due to the presence of dust.  Their intrinsic (unobscured)
colors are similar to those of Lyman break galaxies, but they are
driven toward redder $U-V$ and redder $V-J$ colors along a dust vector
whose slope depends on the nature and distribution of the dust (see,
e.g., Wuyts et al. 2009).  Finally, a population of galaxies with red
$U-V$ colors is present at $z
\sim 2$ whose SED is well matched by that of a passive or quiescently
star-forming galaxy at an older age.  Their $V-J$ colors are relatively
blue compared to those of dusty starbursts at the same optical color.

Labb\'{e} et al. (in preparation) designed a color criterion to select the
quiescent red galaxies based on their rest-frame $U$, $V$, and $J$
photometry.  The selection window is defined as follows:

\begin {equation}
U-V \ge 1.3\ \&\ V-J \le 1.6\ \&\ U-V \ge 0.9 (V-J) + X
\label{colsel.eq}
\end {equation}

where X is 0.31 and 0.4 for our low and high redshift bin respectively.

\begin {figure} [t]
\centering
\plotone{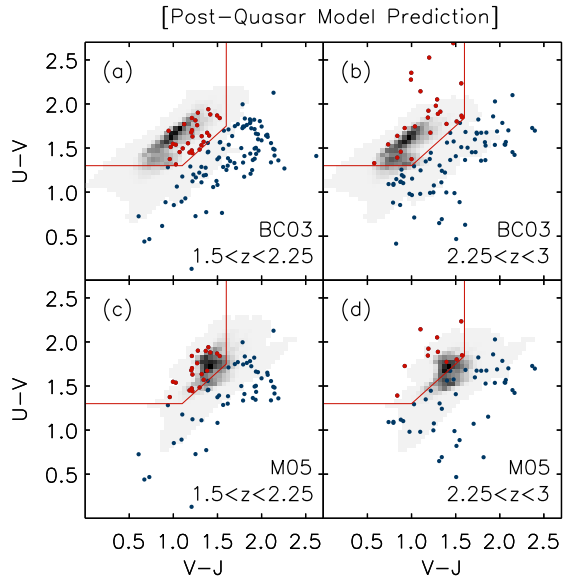} 
\caption{\small Model rest-frame $U-V$ versus $V-J$ color-color
  distribution of simulated galaxies with $\log M > 10.6$ that have
  had a merger and quasar phase in their past ({\it grayscales}), with
  a darker intensity indicating a larger number of post-quasars.
  Observed galaxies above the same mass limit in the FIRES and
  GOODS-South fields are overplotted.  Red symbols mark the galaxies
  that satisfy the quiescent galaxy criterion, the selection window of
  which is marked by the red wedge.  A notable difference between the
  synthetic photometry derived using the BC03 and M05 stellar
  population synthesis code is the redder $V-J$ color in the latter
  case.  Recognizing this uncertainty in the model prediction, we can
  still conclude that the predicted color distribution of post-quasars
  roughly coincides with that of quiescent red galaxies.
\label{UVvsVJpost.fig}}
\end {figure}
The validity of this selection criterion was confirmed by the fact
that galaxies within the UVJ box generally have low MIPS-based
specific star formation rates (Labb\'{e} et al. in prep).  Conversely,
MIPS detected galaxies at $z \sim 2$, suggesting dust-enshrouded star
formation, tend to lie redward of the wedge.  We draw the wedge in
Fig.\ \ref{UVvsVJpost.fig} and indicate the location of all galaxies
with $\log M > 10.6$ in the FIRES and GOODS-South fields in the
color-color diagram.  Red circles mark the objects that satisfy Eq.\
\ref{colsel.eq}.  The upper panels show the observed sample with
BC03-based masses above $\log M > 10.6$, whereas in the bottom panels
the mass limit was applied to the M05-based stellar mass estimates.

We also present a binned representation of the model color-color
distribution of post-quasar galaxies only in Fig.\
\ref{UVvsVJpost.fig}.  The panels correspond to the $1.5<z<2.25$ and
$2.25<z<3$ redshift bins, and model photometry derived from BC03 and
M05 templates respectively.  The color-color distribution computed
with the SMC-like reddening curve from Pei (1992) instead of the
Calzetti et al. (2000) law is not plotted, but looks very similar.  In
the age range of 0.3 - 1 Gyr a dust-free SSP template of M05 has
redder $V-J$ colors than a model of BC03 due to a different
implementation of the thermally pulsing asymptotic giant branch
(TP-AGB) phase (e.g., Maraston et al. 2006).  Summing over all stellar
particles (each treated as a SSP) of the simulated post-starburst
(post-quasar) galaxies, this produces the shift toward redder
integrated $V-J$ colors in panels (c) and (d) with respect to panels
(a) and (b).

In all realizations of the synthetic photometry, the predicted
color-color distribution of the post-quasar population coincides more
or less with the region of color-color space selected by the quiescent
galaxy criterion.  As a control sample, we analyzed a set of disk
simulations, identical in initial conditions to the merger
progenitors, but left to evolve in isolation instead of undergoing a
violent merger process.  Due to the lack of new gas accretion, these
simulated galaxies also reach phases of low specific star formation
rate ($SFR/M < 1/t_{\rm Hubble}$).  However, the vast majority of
these evolved disk galaxies have synthetic colors that place them
outside the quiescent galaxy selection window, in a region of
color-color space where only actively star-forming galaxies are found
in the observations.  We conclude that an abrupt quenching of the star
formation is required to reproduce the properties of observed
quiescent galaxies at high redshift.  Dissipative mergers involving
AGN feedback seem to be a promising mechanism.

\subsubsection {Star-forming galaxies}
\label{UVvsVJSF.sec}

\begin {figure} [t]
\centering
\plotone{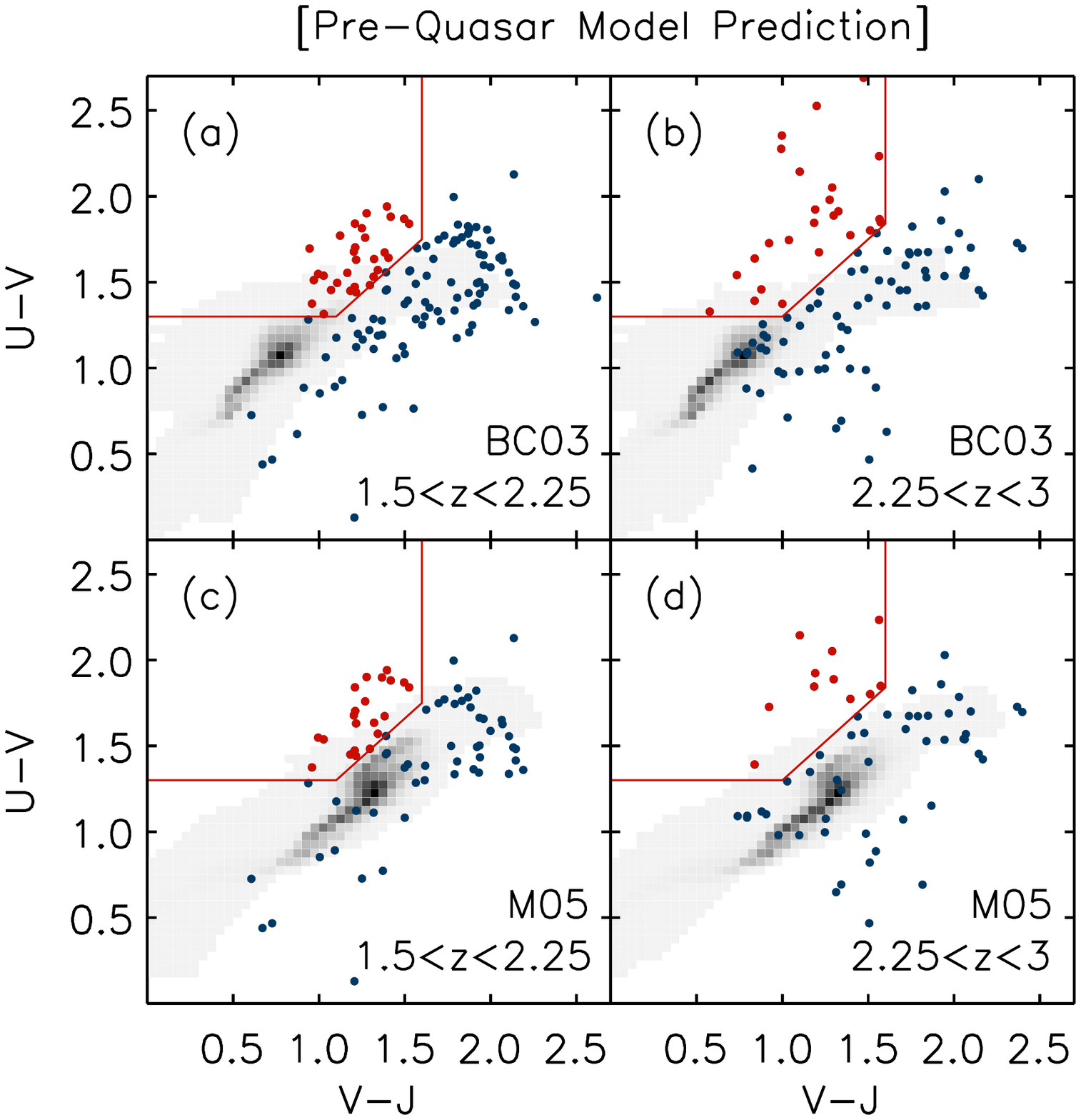} 
\caption{\small Model rest-frame $U-V$ versus $V-J$ color-color
  distribution of simulated galaxies with $\log M > 10.6$ that will
  undergo a quasar phase in less than 700 Myr ({\it grayscales}), with
  a darker intensity indicating a larger number of galaxies.  Observed
  galaxies above the same mass limit in the FIRES and GOODS-South
  fields are overplotted.  Blue symbols mark the galaxies that fall
  outside the quiescent galaxy criterion ({\it red wedge}).  A notable
  difference between the synthetic photometry derived using the BC03
  and M05 stellar population synthesis code is the redder $V-J$ color
  in the latter case.  (a) and (b) The model colors based on BC03 are
  a poor match to the observed star-forming galaxies ({\it blue
  symbols}).  The $V-J$ colors fall blueward of the observed
  distribution, and only the lower half of the observed $U-V$
  distribution of star-forming galaxies is reproduced.  (c) and (d)
  The model colors based on M05 give a better match in the blue $U-V$
  regime, but simulated objects with $V-J \simgeq 1.6$ are nearly
  absent.
\label{UVvsVJpre.fig}}
\end {figure}
A significant fraction ($\sim$ two thirds) of the observed massive galaxy
population at $1.5<z<3$ has colors located outside the quiescent red
galaxy wedge.  These objects reach from blue $U-V$ colors typical for
Lyman break galaxies, which are known to host relatively unobscured
star formation, up to the redder optical and optical-to-NIR colors
from galaxies that are believed to host heavily obscured star
formation.  Here, we investigate whether the predicted color-color
distribution for merging galaxies that will undergo a quasar phase in
less than 700 Myr can reproduce the color range of observed
star-forming galaxies.  Fig.\ \ref{UVvsVJpre.fig} compares the model
prediction ({\it grayscales}) to the observed massive galaxy colors
({\it blue circles} for star-forming galaxies).

As could be anticipated from \S\ref{VJ.sec}, the model photometry does
not reproduce the colors of observed dusty star-forming galaxies ($U-V
> 1.3$ and outside the quiescent red galaxy wedge).  We note that our
control sample of disk galaxies evolving in isolation do not reach the
red optical-to-NIR colors of observed dusty starbursts either during
their actively star-forming lifetime.

At bluer $U-V$, the synthetic photometry based on M05 templates gives a
decent match to the observations, whereas the BC03 colors in
combination with a Calzetti et al. (2000) attenuation law are offset
by a few tenths of a magnitude toward bluer $V-J$.

\section {Specific star formation rate as a function of stellar mass}
\label{sSFR.sec}

So far, we have compared the synthetic colors of merging and
post-quasar galaxies with those of observed star-forming and quiescent
galaxies respectively.  The separation between star-forming and
quiescent galaxies for our observed galaxies was based on their
location in the UVJ diagnostic diagram.  As an independent check, we
now use the UV + 24 $\mu$m derived star formation rates to compare the
observed distribution of specific star formation rates as a function
of stellar mass with the distribution predicted by the merger model.
The specific star formation rate is defined as the ratio of the star
formation rate over the stellar mass.  It equals the inverse of a
mass-doubling time in the case of constant star formation.  Here, we
limit our sample to the GOODS-South field, where the 24 $\mu$m imaging
is sufficiently deep (20 $\mu$Jy, 5$\sigma$) to obtain useful
constraints on the star formation rates.

\begin {figure*} [t]
\centering
\epsscale{0.48}
\plotone{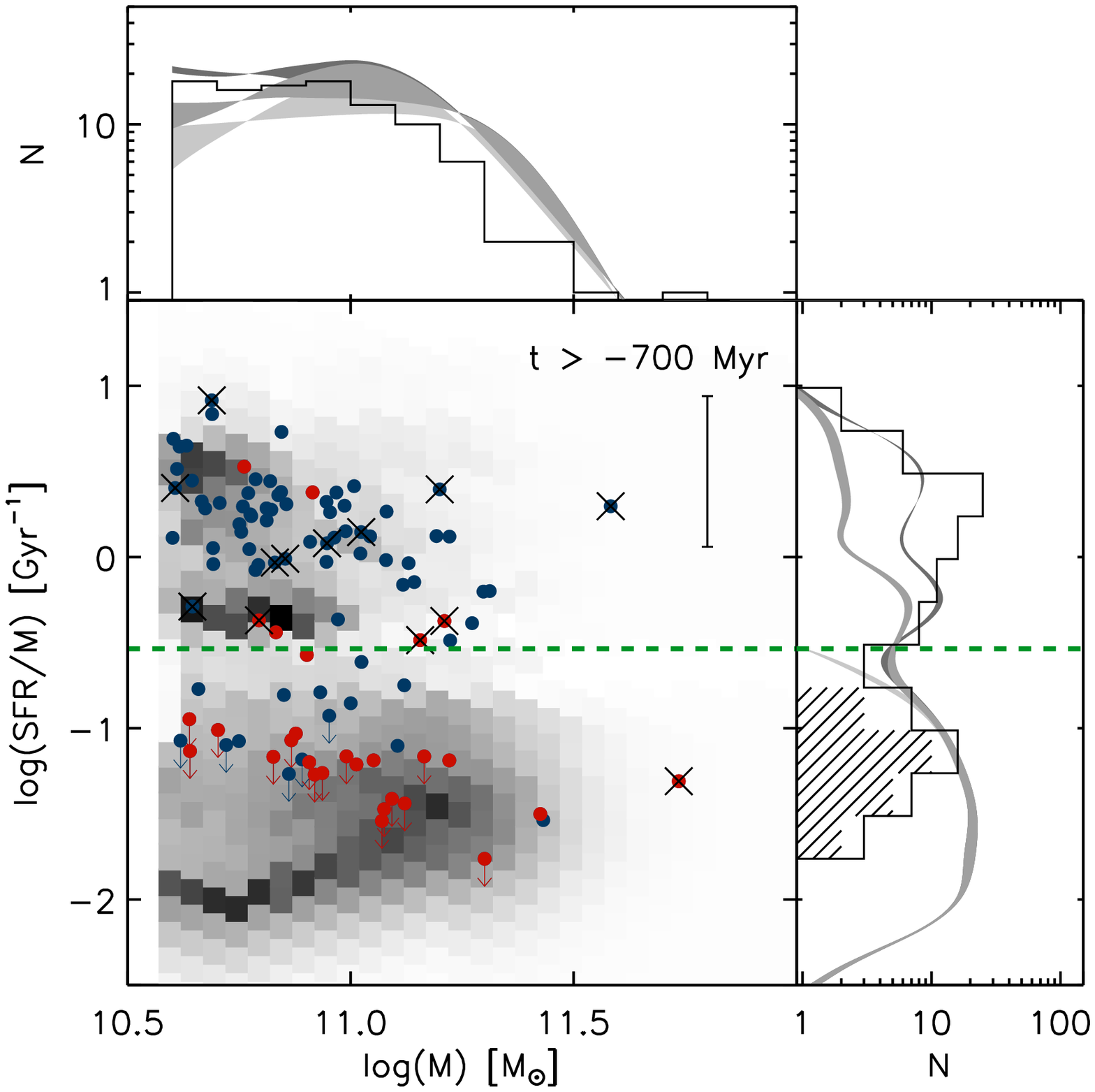}
\plotone{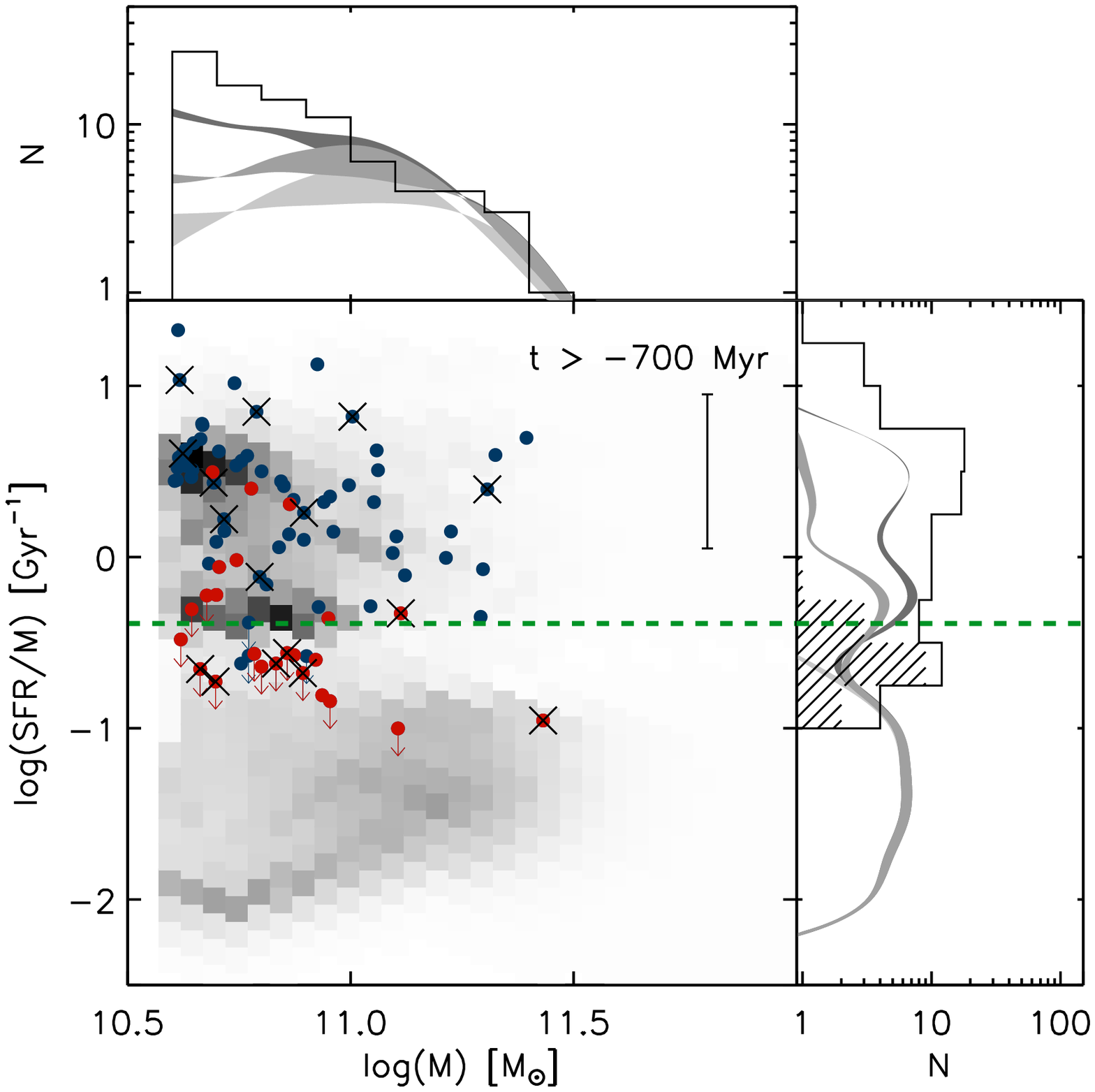} 
\epsscale{1}
\caption{Specific star formation rate as a function of stellar mass
for massive galaxies at $1.5<z<3$ in the GOODS-South field with colors
falling inside ({\it red circles}) or outside ({\it blue circles}) the
selection window for quiescent red galaxies.  Cross symbols indicate
which sources are detected in X-rays.  The vertical error bar
indicates the systematic error in $SFR/M$.  The dashed green line
marks $1/t_{\rm Hubble}$.  The model predictions are plotted with
grayscales.  The top and side panels show the mass and $SFR/M$
distribution, with the black histogram representing the observed
sample, and the grayscaled curves showing the model predictions for
post-quasars and merging galaxies up to 700, 200, and 0 Myr before the
quasar phase.  When integrating down to 700 Myr before the quasar
phase, the predicted number density of galaxies with $SFR/M > 1/t_{\rm
Hubble}$ is 2-3 times smaller than observed, possibly (at least in
part) due to AGN contribution to the 24 $\mu$m emission from which the
observed SFR were derived.
\label{sSFR.fig}}
\end {figure*}
Fig.\ \ref{sSFR.fig} shows the binned model distribution in grayscales
and overplotted are the observed massive galaxies that fall inside
({\it red circles}) and outside ({\it blue circles}) the quiescent red
galaxy wedge.  Upper limits are drawn for objects that were undetected
by MIPS.  Cross symbols mark those objects that are detected in the 1
Ms Chandra X-ray exposure (Giacconi et al. 2002).  We caution that the
24 $\mu$m flux of these objects could have an AGN contribution.
Moreover, Daddi et al. (2007b) recently found that a significant
fraction (20-30\% to $K^{tot}_{Vega}<22$, and up to $\sim 50 - 60\%$
for $M \sim 10^{11}\ M_{\sun}$) of star-forming galaxies that are not
individually detected in the X-rays shows evidence for heavily
obscured AGN by the presence of a mid-IR flux excess.  The vertical
error bar indicates a conservative measure of the systematic
uncertainty in the conversion from 24 $\mu$m flux to the obscured part
of the star formation rate.  The top and side panels show the
distribution of masses and specific star formation rates separately.
With lighter polygons, we illustrate how the predicted distribution
changes when integrating the merger rate function only to the
evaluated redshift or 200 Myr past the evaluated redshift.  The latter
case includes the nuclear starburst phase, but not earlier
star-forming phases.

At $1.5<z<2.25$, the broad-band color criterion is efficient in
distinguishing observed quiescent galaxies from star-forming galaxies
with high specific star formation rates.  In the higher redshift bin,
we are more limited by upper limits on the 24 $\mu$m flux.  The bulk
of broad-band selected quiescent galaxies at $1.5<z<3$ shows smaller
specific star formation rates than their counterparts outside the
broad-band selection window, although some reach values above
$1/t_{\rm Hubble}$ (dashed green line).  The latter objects would
double their stellar mass in less than a Hubble time if they form
stars at a constant rate.  Perhaps they are scattered into the UVJ box
by photometric redshift uncertainties, or their SFR is overestimated.

The model $SFR/M$ distribution is composed of merger-triggered
star-forming galaxies with $SFR/M > 1/t_{\rm Hubble}$, and post-quasar
systems with $SFR/M < 1/t_{\rm Hubble}$.  The precise shape of the
distribution at the low $SFR/M$ end depends on, e.g., the relative
orientation with which the disk progenitors were merged.  However, the
large difference between the star formation mode before and after the
quasar phase is a robust feature of all considered simulations.  The
depth of the MIPS observations inhibits strong observational
constraints on the distribution of individual sources at the low
$SFR/M$ end, but a stacking analysis by Labb\'{e} et al. (in prep)
reveals similarly low $SFR/M$ values ($\sim 3 \times 10^{-2}$
Gyr$^{-1}$) as for simulated post-quasar galaxies.  As in the
observations, in particular at $1.5<z<2.25$, there is a slight hint
that the most heavily star-forming objects reside primarily at the
lower masses within our mass-limited sample.  Papovich et al. (2006)
and Reddy et al. (2006) find that the specific star formation rate is
inversely proportional to mass, implying that the ongoing star
formation at $z
\sim 2$ contributes more significantly to the mass buildup of low-mass
galaxies than to high-mass galaxies.

The predicted abundance of merger-triggered nuclear starbursts,
occuring between 0 and 200 Myr before the quasar phase, seems to be
insufficient to account for all observed massive galaxies with high
specific star formation rates ($SFR/M > 1/t_{\rm Hubble}$).  When
including earlier phases of star formation induced by the merging
event (up to 700 Myr before the quasar phase), we find that the
observed number density of galaxies with $SFR/M > 1/t_{\rm Hubble}$ is
twice as large as predicted by the model.  Part of this offset may be
due to possible AGN contributions to the 24 $\mu$m emission from which
the star formation rates were derived (see, e.g., Daddi et al. 2007b).
Alternatively, this might indicate that not all star-forming galaxies
can be accounted for by the considered merger/quasar driven model for
galaxy evolution.

\section {The number and mass density of massive galaxies at $1.5<z<3$: analysis by type}
\label{analysis_nrho.sec}

We now proceed to quantify the observed and modeled number and mass
densities of different types of massive galaxies at $1.5<z<3$.  As
before, the model prediction was derived by integrating the merger
rate function to include all galaxies that once contributed to the
observed quasar luminosity function or will do so in less than 700
Myr.  From this, we extracted 6 samples using the criteria discussed in
\S\ref{UVvsVJ.sec} and \S\ref{sSFR.sec}.  Four are based on the UVJ diagnostic diagram: 
galaxies above $\log M > 10.6$ (or $\log M > 11$) with broad-band colors satisfying
the quiescent red galaxy criterion (Eq.\
\ref{colsel.eq}, \S\ref{quiescent.sec}), and galaxies above $\log M >
10.6$ (or $\log M > 11$) that do not satisfy Eq.\ \ref{colsel.eq}
(\S\ref{SF.sec}).  Furthermore, we use the MIPS-based specific star
formation rates to independently address the abundance of relatively
quiescent ($SFR/M < 1/t_{\rm Hubble}$, \S\ref{quiescent.sec}) and actively
star-forming ($SFR/M > 1/t_{\rm Hubble}$, \S\ref{SF.sec}) galaxies above
$\log M = 10.6$.  The $\log M > 11$ samples allow us to include the
larger but shallower MUSYC survey in the comparison.  In each case, we
impose an identical selection criterion on the observed sample of
galaxies.

\subsection {The number and mass density of massive quiescent red galaxies}
\label{quiescent.sec}

Having established the similarity in colors of the model post-quasar
population and the observed quiescent red galaxy population above the same mass
limit, we now turn to a comparison of their number and mass densities.
Our aim is to constrain the fraction (in number and mass) of massive
quiescent red galaxies at redshifts $1.5<z<3$ that descendants of
merger-triggered quasars can account for.

\begin {figure*} [htbp]
\centering
\epsscale{0.48}
\plotone{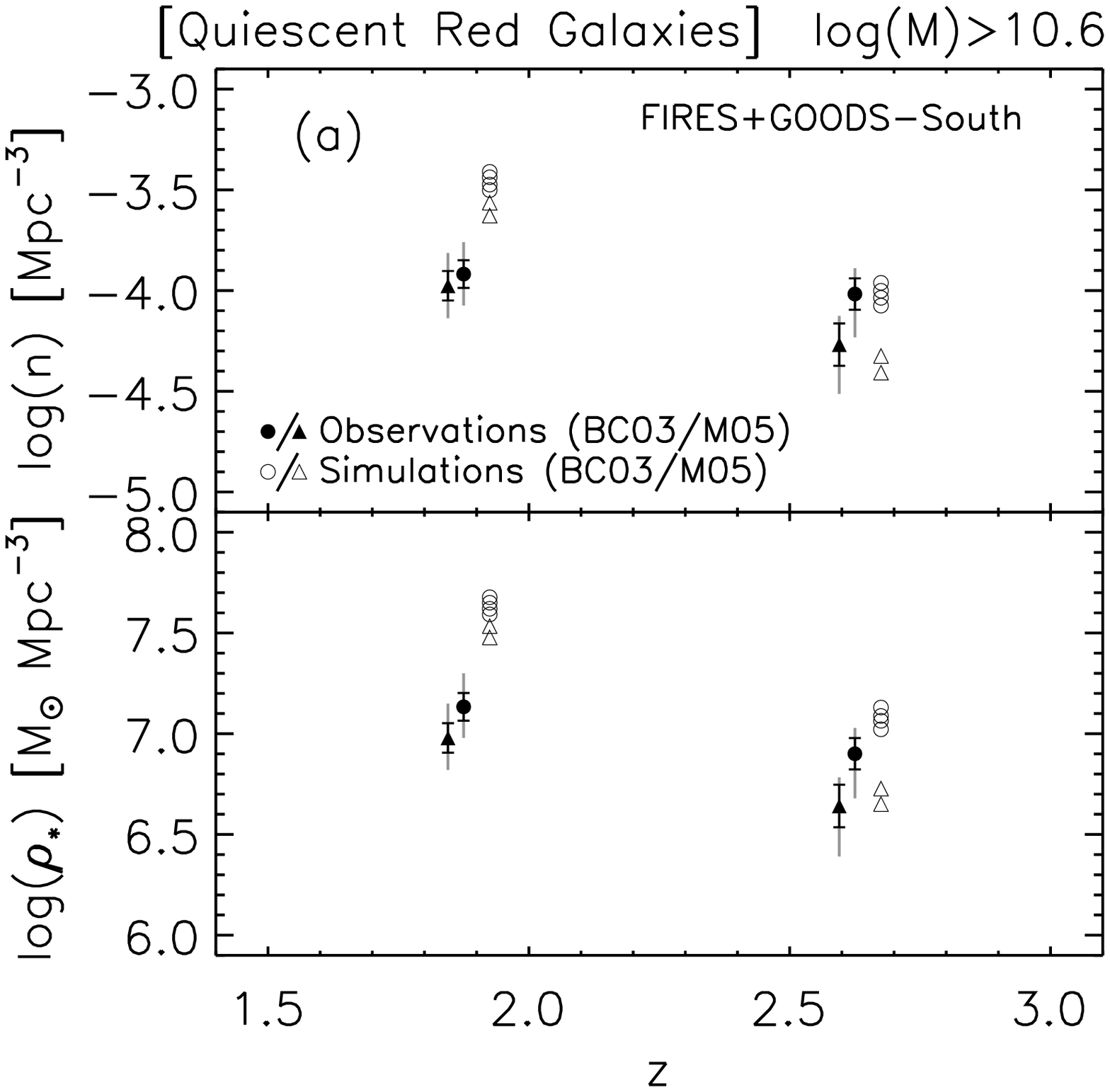}
\plotone{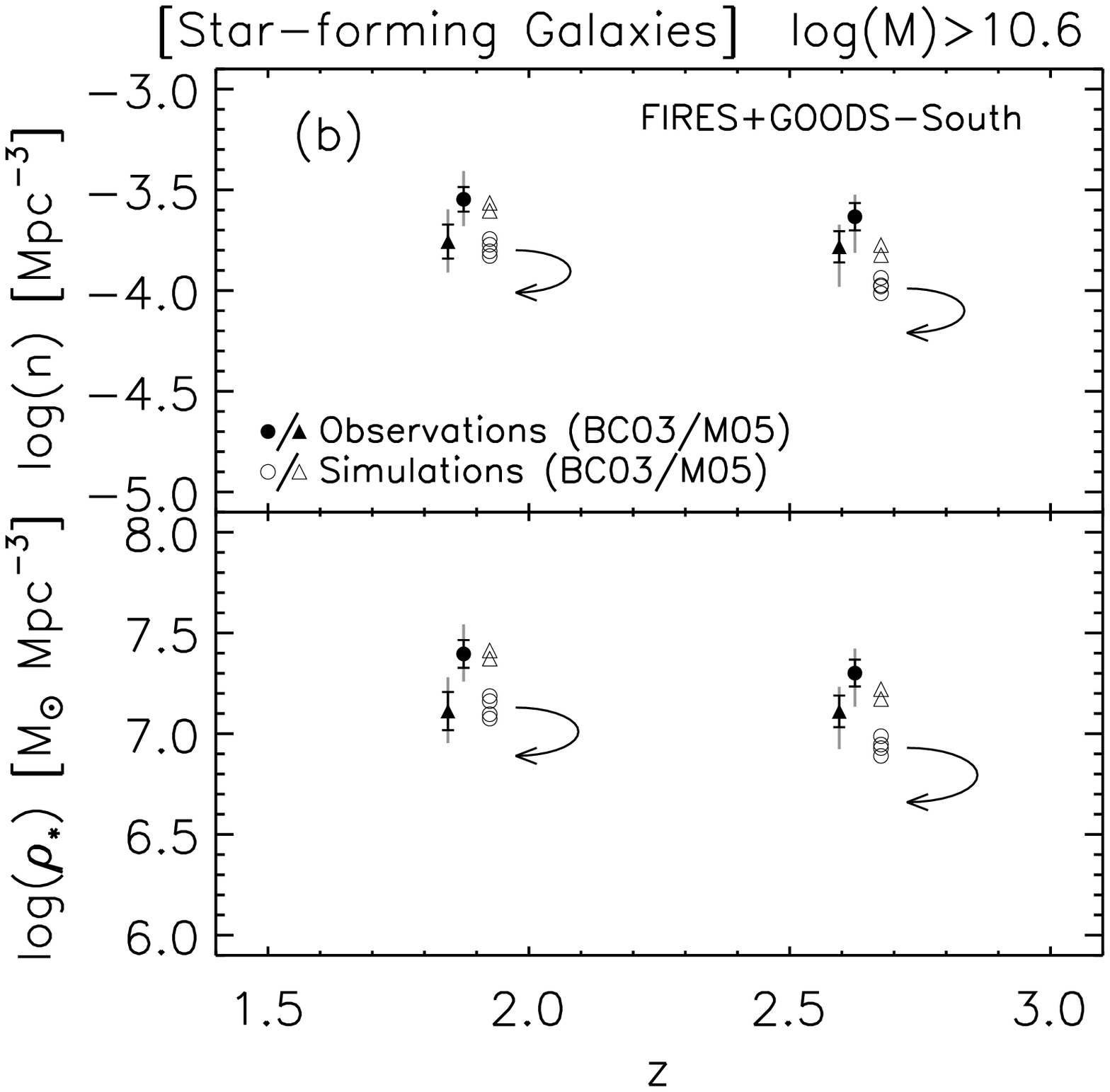} 
\plotone{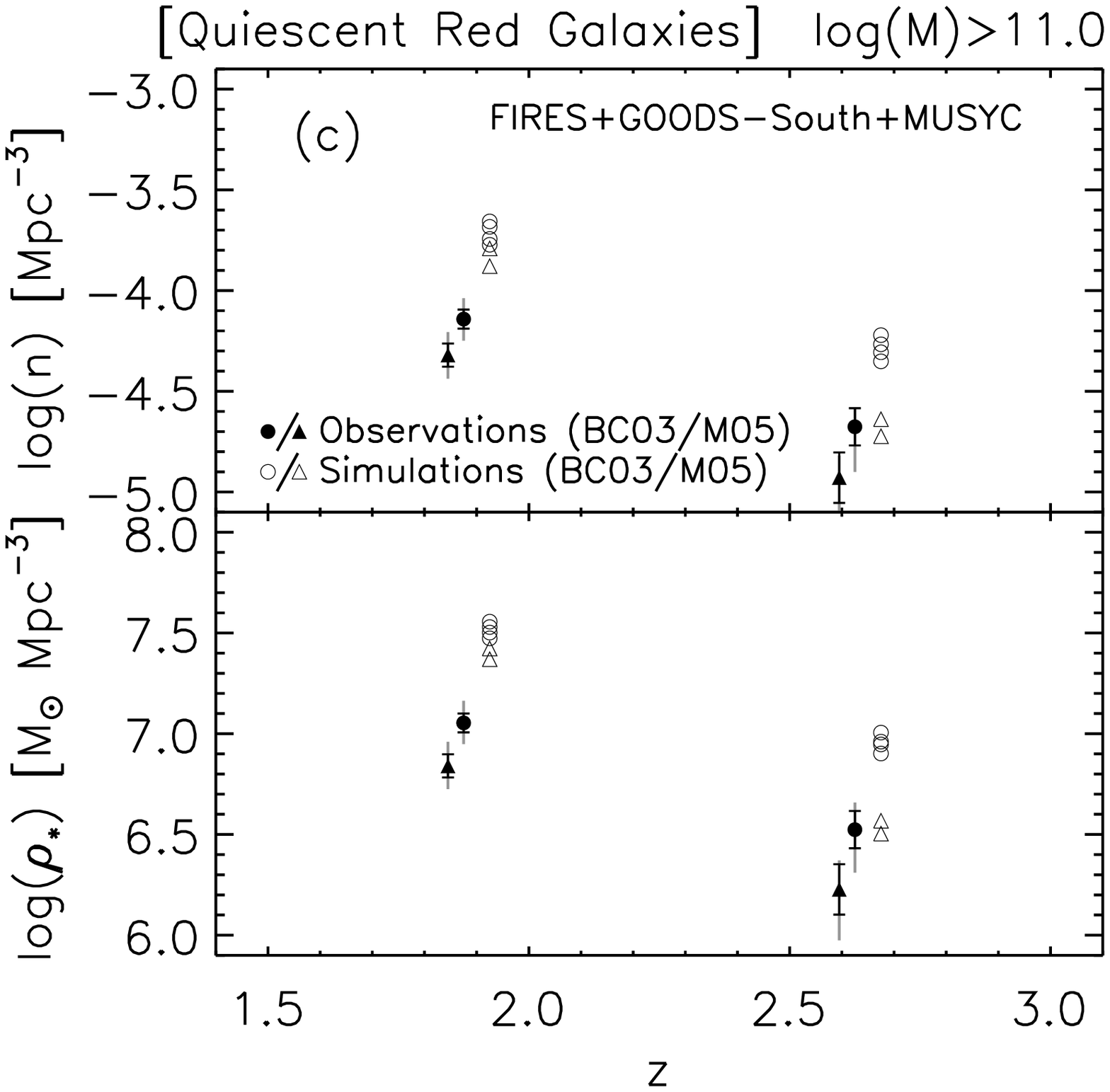}
\plotone{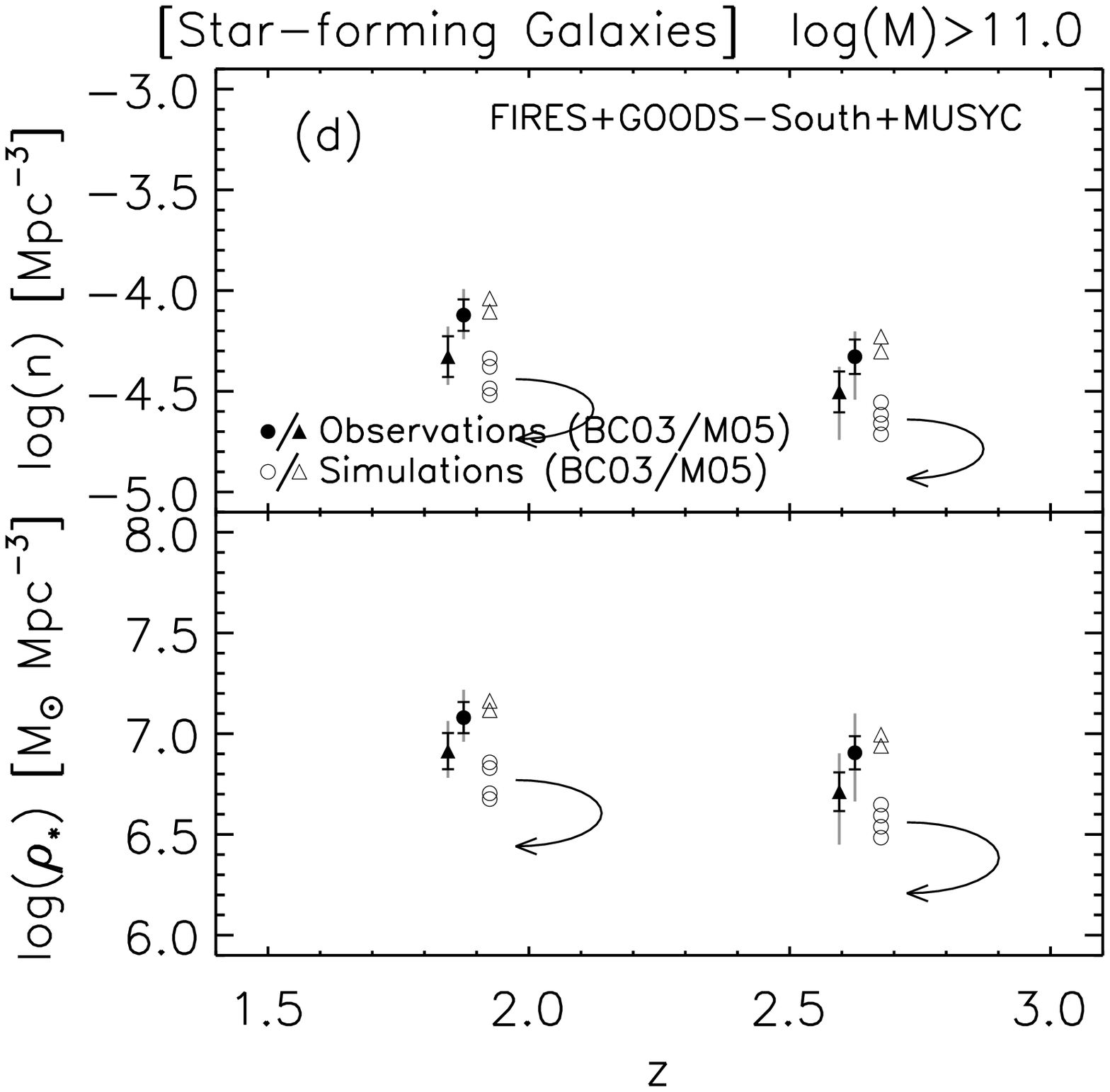} 
\plotone{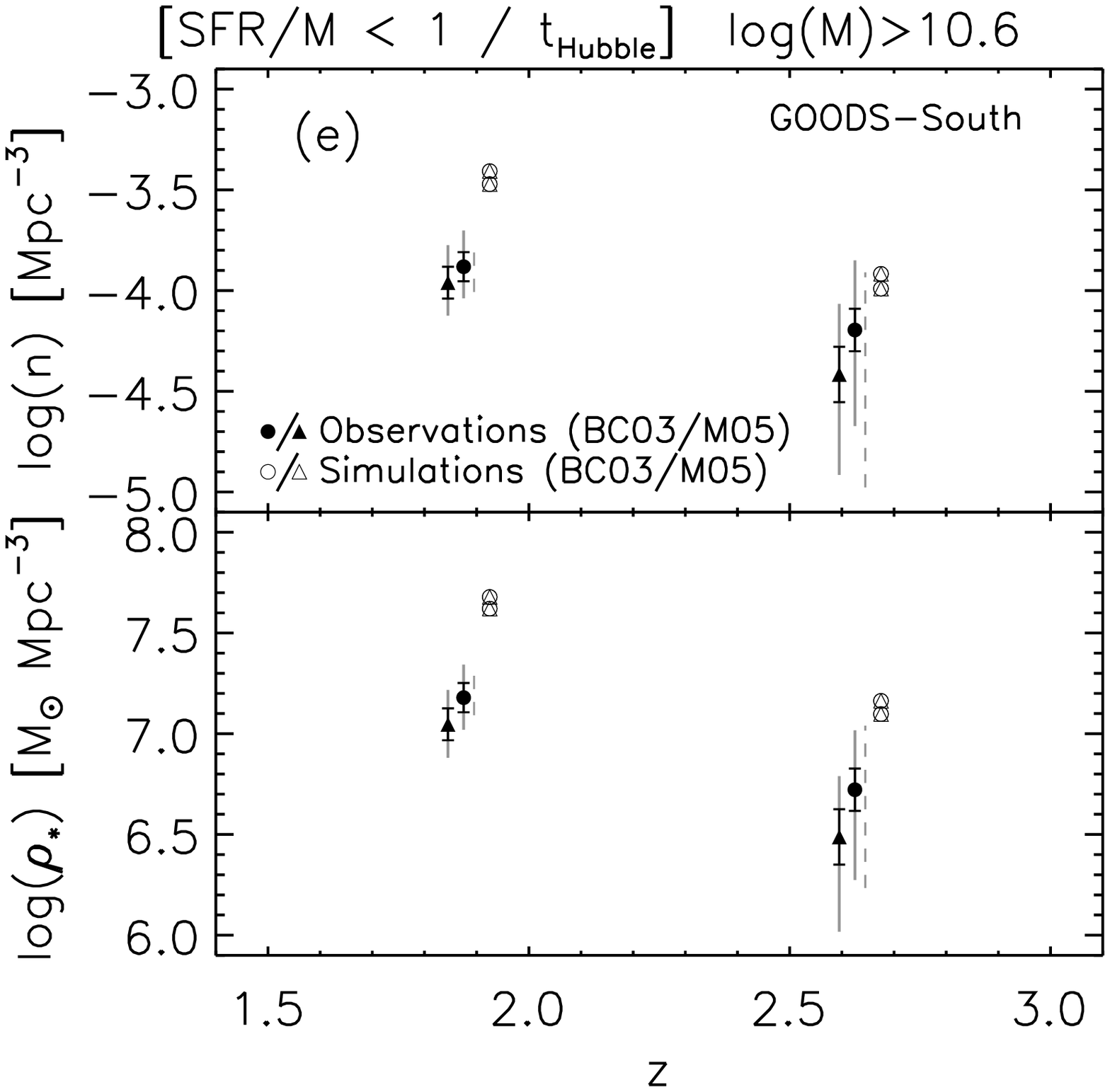}
\plotone{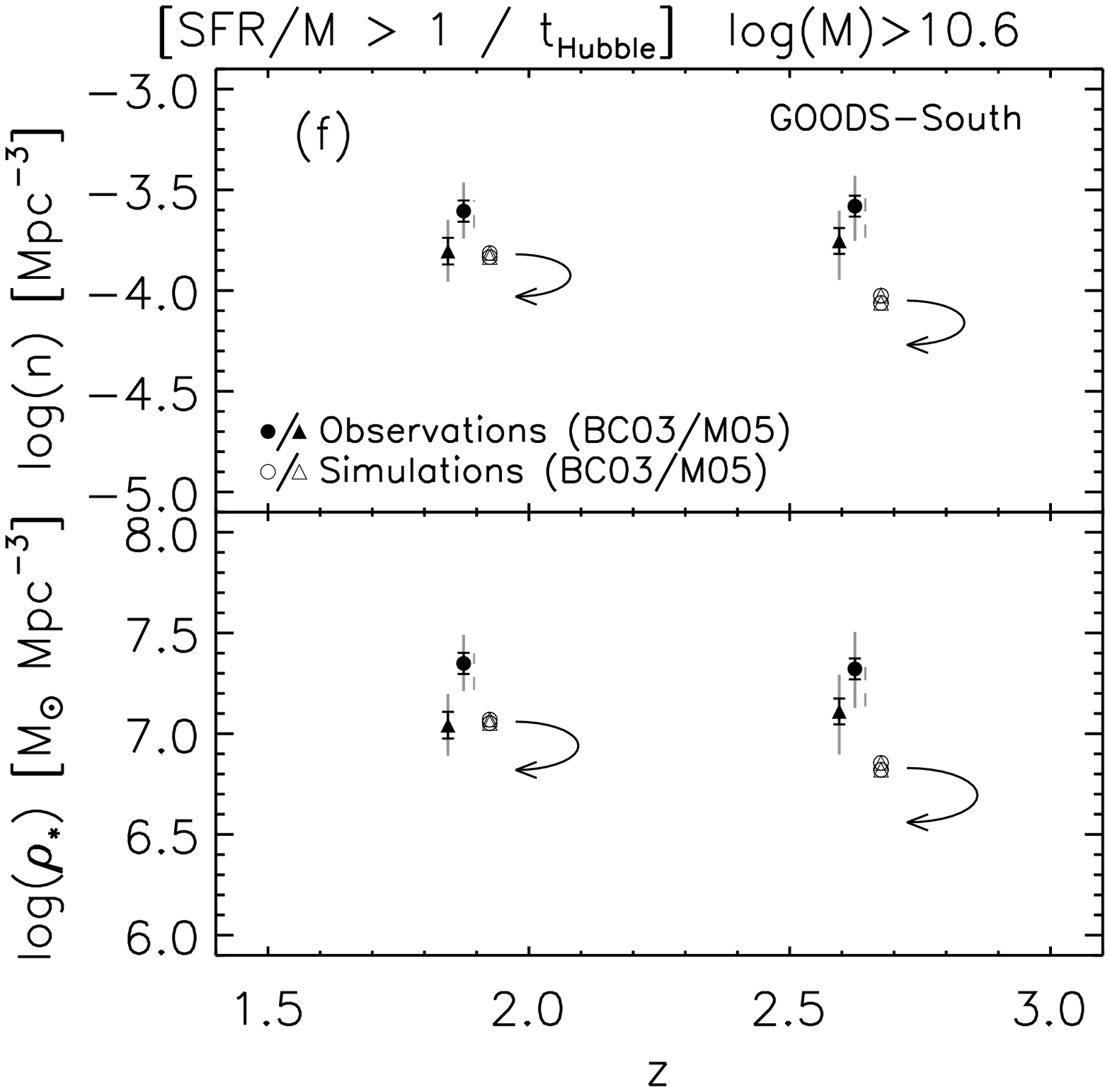} 
\epsscale{1}
\caption{\footnotesize The number and mass density of observed ({\it filled
symbols}) and modeled ({\it empty symbols}) massive galaxies as a
function of redshift above the same mass limit and satisfying the same
selection criterion.  The black error bar represents the Poisson shot
noise solely.  The gray error bar accounts for uncertainties in
redshift, mass, and rest-frame colors and a (mostly dominating)
contribution from cosmic variance.  The dashed error bars in panel (e)
and (f) reflect the systematic uncertainty in the SFR.  Arrows in
panel (b), (d), and (f) indicate how the model results change when
including the effect of mass underestimates in the SED modeling of
star-forming galaxies.  We consider 6 samples: (a) Quiescent red
galaxies with $\log M > 10.6$ in FIRES+GOODS-S, (b) Star-forming
(non-quiescent) galaxies with $\log M > 10.6$ in FIRES+GOODS-S, (c)
quiescent red galaxies with $\log M > 11$ in FIRES+GOODS-S+MUSYC, (d)
star-forming galaxies with $\log M > 11$ in FIRES+GOODS-S+MUSYC, and
finally galaxies with (e) $SFR/M < 1/t_{\rm Hubble}$ or (f) $SFR/M >
1/t_{\rm Hubble}$ and $\log M > 10.6$ in GOODS-S.  We find that the
model tends to overpredict the number of quiescent and underpredict
the number of actively star-forming galaxies by at most a factor 3.
\label{nrho.fig}}
\end {figure*}
\begin{deluxetable*}{lrrrrrrr}
\tablecolumns{8}
\tablewidth{0pc}
\tablecaption{Number and mass densities for massive galaxies \label{nrho.tab}
}
\tablehead{
 & & & \multicolumn{2}{c}{Observations\tablenotemark{a}} & & \multicolumn{2}{c}{Model Prediction\tablenotemark{b}} \\
\cline{4-5} \cline{7-8} \\
\colhead{Type} & \colhead{Mass limit} & \colhead{Redshift} & \colhead{n} & \colhead{$\rho_*$} & & \colhead{n} & \colhead{$\rho_*$} \\
 & $M_{\sun}$ & & $10^{-4}\ Mpc^{-3}$ & $10^7\ M_{\sun}\ Mpc^{-3}$ & & $10^{-4}\ Mpc^{-3}$ & $10^7\ M_{\sun}\ Mpc^{-3}$ \\
}
\startdata
All                   & $4 \times 10^{10}$ & $1.5<z<2.25$ & $4.0^{+1.3}_{-1.3}$ & $3.8^{+1.3}_{-1.2}$ & & $4.8 - 5.5$ & $5.4 - 6.0$ \\
All                   & $4 \times 10^{10}$ & $2.25<z<3$   & $3.3^{+1.1}_{-1.2}$ & $2.8^{+0.9}_{-0.9}$ & & $1.9 - 2.2$ & $1.9 - 2.2$ \\
Quiescent             & $4 \times 10^{10}$ & $1.5<z<2.25$ & $1.2^{+0.5}_{-0.5}$ & $1.4^{+0.6}_{-0.6}$ & & $2.4 - 3.9$ & $3.0 - 4.8$ \\
Quiescent             & $4 \times 10^{10}$ & $2.25<z<3$   & $1.0^{+0.4}_{-0.5}$ & $0.8^{+0.3}_{-0.4}$ & & $0.4 - 1.1$ & $0.4 - 1.3$ \\
Star-forming          & $4 \times 10^{10}$ & $1.5<z<2.25$ & $2.8^{+1.1}_{-1.0}$ & $2.5^{+1.0}_{-0.9}$ & & $1.5 - 2.7$ & $1.2 - 2.6$ \\
Star-forming          & $4 \times 10^{10}$ & $2.25<z<3$   & $2.3^{+0.9}_{-1.0}$ & $2.0^{+0.8}_{-0.8}$ & & $1.0 - 1.7$ & $0.8 - 1.7$ \\
Quiescent             & $10^{11}$          & $1.5<z<2.25$ & $0.7^{+0.2}_{-0.2}$ & $1.1^{+0.3}_{-0.3}$ & & $1.3 - 2.2$ & $2.3 - 3.6$ \\
Quiescent             & $10^{11}$          & $2.25<z<3$   & $0.2^{+0.1}_{-0.1}$ & $0.3^{+0.2}_{-0.1}$ & & $0.2 - 0.6$ & $0.3 - 1.0$ \\
Star-forming          & $10^{11}$          & $1.5<z<2.25$ & $0.8^{+0.3}_{-0.2}$ & $1.2^{+0.5}_{-0.4}$ & & $0.3 - 0.9$ & $0.5 - 1.5$ \\
Star-forming          & $10^{11}$          & $2.25<z<3$   & $0.5^{+0.3}_{-0.2}$ & $0.8^{+0.7}_{-0.4}$ & & $0.2 - 0.6$ & $0.3 - 1.0$ \\
$SFR/M < 1/t_{\rm Hubble}$ & $4 \times 10^{10}$ & $1.5<z<2.25$ & $1.3^{+0.7}_{-0.6}$ & $1.5^{+0.7}_{-0.7}$ & & $3.4 - 3.9$ & $4.2 - 4.8$ \\
$SFR/M < 1/t_{\rm Hubble}$ & $4 \times 10^{10}$ & $2.25<z<3$   & $0.6^{+0.9}_{-1.2}$ & $0.5^{+0.6}_{-0.9}$ & & $1.0 - 1.2$ & $1.3 - 1.5$ \\
$SFR/M > 1/t_{\rm Hubble}$ & $4 \times 10^{10}$ & $1.5<z<2.25$ & $2.5^{+1.0}_{-0.9}$ & $2.2^{+0.9}_{-0.8}$ & & $1.5 - 1.5$ & $1.1 - 1.2$ \\
$SFR/M > 1/t_{\rm Hubble}$ & $4 \times 10^{10}$ & $2.25<z<3$   & $2.6^{+1.3}_{-1.0}$ & $2.1^{+1.3}_{-1.0}$ & & $0.9 - 0.9$ & $0.7 - 0.7$ \\
\enddata
\tablenotetext{a}{\scriptsize The error bars in the observed densities account for Poisson noise, cosmic variance, and the uncertainties in redshift, rest-frame color and mass of the individual galaxies.  They do not account for the systematic dependence on the stellar population synthesis code used to derive the stellar masses (values given here are for BC03), nor was the systematic uncertainty in the conversion from 24 $\mu$m to SFR (of the order of 1 dex) included in the results for the sample selected on $SFR/M$.}
\tablenotetext{b}{\scriptsize The range in model densities indicates a crude estimate of the size of uncertainties in the merger rate function and the dependence on choice of attenuation law and stellar population synthesis code to compute the synthetic photometry.}
\end{deluxetable*}
In order to do this, we selected the observed and modeled galaxies
with $\log M > 10.6$ that lie inside the wedge defined by Eq.\
\ref{colsel.eq} and compute the number and mass density for the probed
comoving volume of $\sim 3.5 \times 10^{5}\ Mpc^3$ in each redshift
bin.  The resulting number and mass densities are plotted as a
function of central redshift of the redshift bin in Fig.\
\ref{nrho.fig}(a).  The filled circles and triangles indicate the observed number
and mass density of quiescent red galaxies above $\log M = 10.6$ using
BC03- and M05-based stellar masses respectively.  Their values and corresponding
uncertainties are listed in Table\ \ref{nrho.tab}.  As in
\S\ref{all_density.sec}, the black error bars account for Poisson shot
noise.  The gray error bars also include selection uncertainties
stemming from uncertainties in the redshift, mass, and rest-frame
colors of individual galaxies, and a dominating contribution from
cosmic variance.  When correcting to the same IMF and mass limit, our
estimates of the number density are in good agreement with those of
smaller samples of galaxies by Cimatti et al. (2008) and Kriek et
al. (2008) that are spectroscopically confirmed to be quiescent.

The empty symbols on Fig.\ \ref{nrho.fig}(a) indicate the predicted
number and mass density of galaxies with $\log M > 10.6$ at
$1.5<z<2.25$ and $2.25<z<3$ whose synthetic photometry places them
within the selection wedge for quiescent red galaxies.  95\% of these
modeled galaxies are in a post-quasar phase of their evolution.  The
different empty symbols represent predictions derived with the BC03
and M05 stellar population synthesis codes, with the Calzetti et
al. (2000) attenuation law and the SMC-like attenuation law from Pei
(1992).  Their spread gives a crude indication of the freedom allowed
by the model.  It also takes into account the uncertainty in the
merger rate function used to populate our model universe with the
binary merger simulations.  As noted already in
\S\ref{UVvsVJdead.sec}, synthetic photometry of post-quasar galaxies
based on M05 templates places part of them at redder $V-J$ than the
diagonal of the UVJ box, even though their sSFRs are low ($10^{-2} -
10^{-0.5}$ Gyr$^{-1}$).  This explains why the M05-based model
predictions in Fig.\ref{nrho.fig}(a) are lower than those based on our
default BC03 models.  We therefore consider the BC03-based model
predictions of the quiescent galaxy abundance as the most reliable.

We find that the model predicts a number density of quiescent galaxies
at $z \sim 1.9$ that is 2.5 times larger and a mass density that is 3
times larger than observed.  At $z \sim 2.6$, the model and
observations are consistent within the error bars.  In other words, if
anything, assuming a one-to-one correspondence between quasars and
gas-rich mergers, the model by Hopkins et al. (2006b) overpredicts the
abundance of merger remnants (i.e., post-quasar galaxies) at
$1.5<z<3$.  The model predicts an increase by a factor 3.5 in the
number and mass density for massive post-quasar galaxies in the 1 Gyr
that passed between $z=2.6$ and $z=1.9$.  The observed sample seems to
suggest less evolution (a factor 1.2 in number density and 1.8 in mass
density), and is even formally consistent with a non-evolving number
and mass density over the considered redshift range.

In order to reduce the effect of cosmic variance, we now include the
MUSYC fields in our analysis, increasing the area by a factor 3.6 and
reducing the cosmic variance for a given mass limit with a similar
factor.  However, this goes at the cost of depth: the 90\%
completeness limit for the MUSYC fields is 1 magnitude shallower than
for GOODS-South.  Consequently, we are restricted to a sample limited
at $M > 10^{11}\ M_{\sun}$, even then requiring a 19\% correction for
incompleteness in the $2.25<z<3$ bin.  Since cosmic variance is larger
for more massive galaxies, the reduction of uncertainties due to
cosmic variance with respect to the deeper FIRES+GOODS-S sample is
more modest than the increase in area.  Fig.\ \ref{nrho.fig}(c) shows
the number and mass density of $M > 10^{11}\ M_{\sun}$ galaxies that
fall within the quiescent red galaxy wedge, as a function of redshift
for the combined FIRES, FIREWORKS, and MUSYC surveys ({\it filled
black symbols}).  Poisson noise is negligible for this sample.  The
gray error bars again account for cosmic variance and the
uncertainties in redshift, rest-frame color, and mass of the
individual galaxies making up the sample.

The main conclusion drawn from Fig.\ \ref{nrho.fig}(c), based on a
largely independent sample, is consistent with that of Fig.\
\ref{nrho.fig}(a).  Namely, a model in which every observed quasar
produces a massive quiescent galaxy overpredicts the abundance of such
galaxies at $1.5<z<3$, by a factor of 3 for the sample above $10^{11}\
M_{\sun}$.  However, for our $\log M > 11$ sample we do find that the
quiescent population grows by a similar amount (a factor $\sim$ 3.5)
in model and observations between $z \sim 2.6$ and $z \sim 1.9$.

When comparing the abundance of quiescent galaxies selected by their
MIPS-based $SFR/M < 1/t_{\rm Hubble}$ (Fig.\ \ref{nrho.fig}(e)), we
arrive at a similar conclusion.  Here, we restricted ourselves to the
GOODS-S field, for which deep MIPS 24$\mu$m data is available.  If
galaxies form stars at a constant rate, the adopted $SFR/M$ threshold
separates galaxies that double their mass in less than a Hubble time
from those that formed the bulk of their stars prior to the time of
observation.  We note that this threshold coincides with the minimum
in the observed and modeled $SFR/M$ distribution (Fig.\
\ref{sSFR.fig}).  For the high-redshift ($z \sim 2.6$) bin, the
presence of a bimodal $SFR/M$ distribution could not be
observationally confirmed or ruled out due to the larger upper limits
on individual $SFR/M$ measurements.  As in Fig.\ \ref{nrho.fig}(a), we
find the model number and mass density to be larger than observed, by
a factor 2.5 and 3 respectively.

Possibly, our results suggest that part of the quasar-descendants are
rejuvenated, e.g. by new gas infall fueling recurrent star formation.
Large cosmological simulations including a prescription for AGN
feedback, such as undertaken by Di Matteo et al. (2008), but with a
resolution comparable to the binary merger simulations presented here
are necessary to investigate such scenarios.

We should also bear in mind that the evolution of the $M_{\rm BH} -
M_*$ relation and its scatter with redshift is only poorly
constrained.  Since this relation is used to translate observed quasar
demographics into galaxy demographics (see
\S\ref{methodology.sec}), an uncertainty of a factor of a few in
$M_{\rm BH} - M_*$ might resolve the offset in abundance as well.
Finally, we caution that beaming effects at the bright end of the
quasar luminosity function might complicate the methodology described
in \S\ref{methodology.sec}.

\subsection {The number and mass density of massive star-forming galaxies}
\label{SF.sec}

Following identical procedures as outlined above, we analyze the
number and mass density of massive galaxies with colors outside the
quiescent red galaxy wedge in Fig.\ \ref{nrho.fig}(b).  Again, we used
a Monte Carlo simulation to determine how many galaxies moved into or
out of the selection window when perturbing their photometry within
the error bars and hence changing the derived properties such as mass
and rest-frame photometry.

We find massive star-forming galaxies in the observed fields to be 2.3
times more abundant in number and 2 times more in mass than massive
quiescent galaxies at $1.5<z<3$.  Their contribution to the overall
number and mass density decreases slightly, to 60\%, when considering
our $\log M > 11$ sample (Fig.\ \ref{nrho.fig}(d)).  

Given that M05-based synthetic photometry placed some of the simulated
post-quasar galaxies with low sSFRs in the star-forming part of the
UVJ diagram, we focus on the results obtained using BC03 templates
(circles).  The model seems to predict that massive star-forming
systems are less, rather than more, abundant than high-redshift
quiescent systems above the same mass limit.  Whereas the model
overpredicted the quiescent number and mass density by factors of a
few (\S\ref{quiescent.sec}), the abundance of the star-forming
population is underpredicted by a similar amount: a factor 1.8 at $z
\sim 1.9$ and a factor 2.2 at $z \sim 2.6$.

We now compare the abundance of actively star-forming galaxies again,
using a sSFR threshold as selection criterion, rather than the
rest-frame optical to NIR colors.  Selecting galaxies with $SFR/M >
1/t_{\rm Hubble}$ from FIREWORKS, we find an observed number density of $2.5
\times 10^{-4}\ Mpc^{-3}$ and $2.6 \times 10^{-4}\ Mpc^{-3}$ at
$1.5<z<2.25$ and $2.25<z<3$ respectively (Fig.\ \ref{nrho.fig}(f)).
Since we interpreted all the 24 $\mu$m emission as dust re-emission
from star formation, the true number density can be lower depending on
the contribution from AGN (see, e.g., Reddy et al. 2005; Papovich et
al. 2006; Daddi et al. 2007b).  The merger model predicts an abundance
of galaxies with high specific star formation rates that is 1.7 (3.0)
times smaller than observed in our low (high) redshift bin.  Despite
the possible AGN contribution to the 24 $\mu$m emission and the large
systematic uncertainty in the conversion from 24 $\mu$m to the
dust-obscured contribution to the star formation rate ({\it dashed
line} in Fig.\ \ref{nrho.fig}(f)), the different selection of actively
star-forming systems produces a similar result as derived from the UVJ
diagram.

Testing SED modeling on mock observations of hydrodynamic merger
simulations, Wuyts et al. (2009) caution for systematic mass
underestimates during phases of merger-induced star formation.  As
discussed in \S\ref{obsprocedure.sec}, accounting for such an effect would increase
the offset with respect to the observed abundance, such that the
merger model then accounts for a third of the observed star-forming
galaxies with $\log M > 10.6$ at $z \sim 1.9$, and a quarter at $z
\sim 2.6$.  This result may be consistent with recent evidence for 
other mechanisms than major mergers driving part of the star formation
at high redshift (Daddi et al. 2007; Shapiro et al. 2008; Genel et
al. 2008).  We note that the fraction of star-forming galaxies
accounted for by the merger model decreases when only considering the
short-lived ($\sim 100$ Myr) starburst at final coalescence instead of
counting all merger-triggered star-forming phases as we do by
integrating the merger rate function to 700 Myr before the peak of
quasar activity.

Overall, we conclude that the model abundances of massive galaxies are
formally consistent with the observations.  However, dividing the
sample in quiescent and star-forming populations, offsets of factors
2-3 occur, with the model predictions being higher for quiescent and
lower for star-forming galaxies.

\section {Pair statistics}
\label{pairstat.sec}

In our model predictions of the number and mass densities of different
samples of massive galaxies, a simple criterion determines whether we
use the integrated properties (color, mass, SFR) for the merging pair
or treat the two progenitors as resolved systems, counting each one
separately, contributing half the mass and SFR (see \S\
\ref{methodology.sec}).  Evolutionary phases, redshifts and viewing
angles for which the projected angular separation between the central
SMBHs is less than $1\farcs 5$ were considered unresolved.  All
post-quasar predictions are robust against the precise form of the
criterion, since by that time the two progenitors have formed one
galaxy.  For earlier phases, applying the criterion decreases the mass
density of massive galaxies, since galaxies drop out of the
mass-limited sample.  The effect on the number density is less
trivial.  On the one hand, galaxies drop out of the mass-limited
sample.  On the other hand, some merging pairs contribute twice.

Here, we focus on an additional test of the merger model allowed by
the fact that some of the pairs will be resolved into two objects.  If
a significant fraction of the massive galaxy population at $1.5<z<3$ is
indeed related to merging events, as our analysis suggests, we expect
to see an excess in the pair statistics with respect to a random
distribution of galaxies on the sky.

\begin {figure} [t]
\centering \plotone{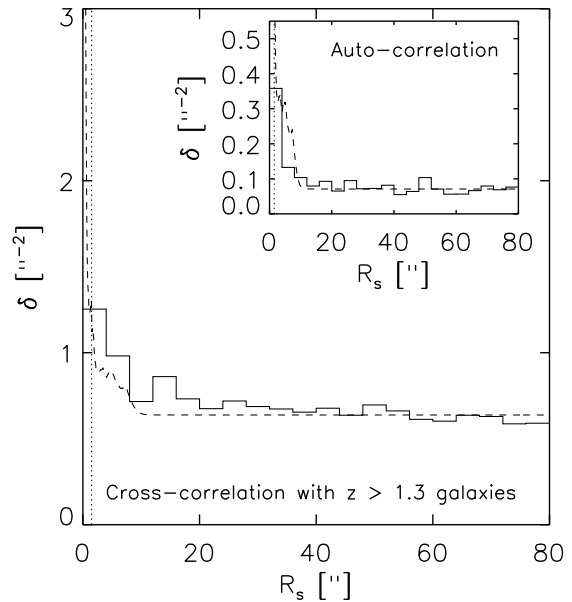} 
\caption{Relative galaxy density ($\delta$) as a function of massive
($\log M > 10.6$) $1.5<z<3$ galaxy to $z>1.3$ galaxy separation
($R_s$) in the GOODS-South field.  The distribution predicted by the
merger model is indicated with the dashed line.  At separations
smaller than $1 \farcs 5$ ({\it dotted line}) an increasing number of
galaxy pairs, if present, will be missed because they would be
detected as a single object.  We find a clear excess at small pair
separations ($R_s < 8''$), as predicted by the merger model.  A weak
pair excess is also visible when only considering the distribution of
separations between massive $1.5<z<3$ galaxies ({\it inset panel}),
but the excess is much below the prediction.
\label{pair.fig}}
\end {figure}
We present the distribution of galaxy-galaxy separations in the
GOODS-South field in Figure\ \ref{pair.fig} ({\it solid histogram}).
We decide not to include the other fields, to prevent differences in
depth from influencing the pair excess signal.  The main panel shows
the results from a cross-correlation of our massive ($\log M > 10.6$)
galaxy sample at $1.5<z<3$ with the sample of all galaxies above
$z>1.3$ in the GOODS-South field, thus avoiding the risk of losing
pair members that by a typical photometric redshift error were placed
at some lower redshift.  For each massive galaxy at $1.5<z<3$, we
measure the distance to all $z>1.3$ galaxies.  We compute the
statistic

\begin {eqnarray}
\delta (R_s) = \frac{\sum\limits_{i=1}^j N_i(R_s)}{\pi \left((R_s + \epsilon)^2 - (R_s - \epsilon)^2\right)}
\end {eqnarray}

where $j$ is the total number of objects in our massive galaxy sample
and $N_i(R_s)$ is the number of $z>1.3$ galaxies that lie between a
distance $R_s - \epsilon$ and $R_s + \epsilon$ from galaxy $i$.  For a
random uniform distribution of galaxies, $\delta (R_s)$ will be flat.
Figure\ \ref{pair.fig} shows that for our sample of massive galaxies
at $1.5<z<3$, this is clearly not the case.  An excess of pairs at
$R_s < 8''$ is visible, also when we consider the distribution of
separations between members of the massive galaxy sample at $1.5<z<3$
only ({\it inset panel}).  We note that, using the wider area UDS
field, Quadri et al. (2008) found an upturn of the correlation
function of distant red galaxies at $2<z<3$ on similarly small scales
($\theta < 10''$).  Spectroscopic relative velocity measurements are
needed to assess what fraction of the small scale excess is due to
bound pairs in the process of merging, and how much can be attributed
to an enhanced number of projected pairs due to clustering.

From the simulations, we measured the physical separations between the
2 merging galaxies and computed the distribution of separation angles
in arcseconds on the sky using the merger rate function (see
\S\ref{methodology.sec}) and assuming random viewing angles.  Adding the mean value of $\delta$ as
measured in the interval $30''<R_s<80''$, we obtain a model prediction
({\it dashed line}) that is in qualitative agreement with the
cross-correlation results, but larger than the weak pair excess seen
in the auto-correlation.  Admittedly, the predicted distribution is
subject to the orbital configuration set at the start of the
simulation, an effect that is not explored in this paper.

\section {Comments and caveats}
\label{comments.sec}

In this Section, we list a number of caveats, and indicate prospects
for improvements on both the model and observational side.  We discuss
aspects affecting the determination of galaxy abundances as well as a
number of possible reasons for the discrepancy between the synthetic
and observed colors of massive star-forming galaxies.  Future
investigations along these lines will help to further test the merger
model.

\subsection {Simulating the observing procedure}
\label{obsprocedure.sec}

\subsubsection {Color gradients}
\label{colorgradients.sec}
First, it is possible that the colors of observed and modeled galaxies
are in fact in agreement, but that a discrepancy was found because we
did not simulate the whole observing procedure.  The observed
optical-to-NIR colors are measured on PSF-matched images within
apertures of size 1'' to 2''.  The IRAC photometry was performed
within apertures of 3'' diameter, and scaled to the smaller color
aperture assuming all sources had a flat $K_s$ - IRAC color profile
(see Wuyts et al. 2008).  The synthetic colors instead were based on
integrated photometry of all stellar particles, irrespective of their
distance to the galaxy center.  The presence of a color gradient with
redder emission in the central regions of the galaxies could therefore
induce an offset in colors in the observed direction.

\begin {figure} [t]
\centering \plotone{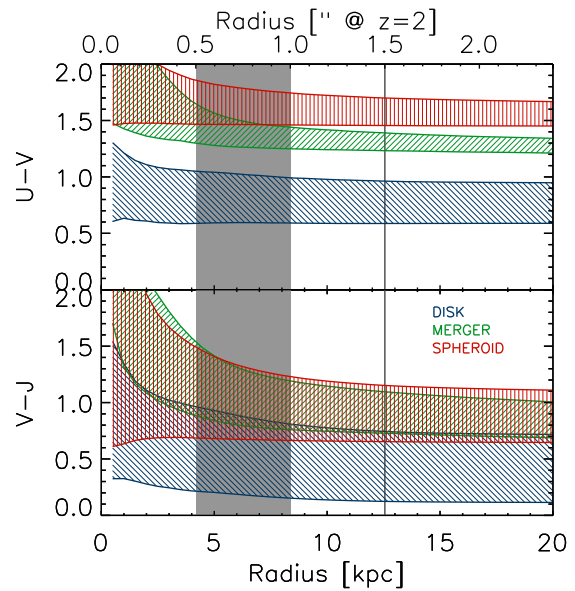} 
\caption{Simulated rest-frame $\rfUV$ and $\rfVJ$ colors before ({\it blue}), during ({\it green}), and after ({\it red}) 
a merger that ends with $M_{*,{\rm final}} = 1.2 \times 10^{11}\
M_{\sun}$ as function of radius in which the color was measured.  The
height of the polygons indicates the range of colors for the same
snapshot, as seen from different angles.  Aperture sizes for the
optical-to-NIR (1'' $<$ diameter $<$ 2'') and IRAC photometry (3''
diameter) of real galaxies are marked in gray.  Color gradients are
present with the galaxy core being redder than the outskirts,
particularly during the merger-driven nuclear starburst ({\it green}).
\label{colgrad.fig}}
\end {figure}
In Fig.\ \ref{colgrad.fig}, we investigate the presence of color
gradients for three snapshots (before, during and after the merger) of
a simulation with final stellar mass $M_{*,{\rm final}} = 1.2 \times 10^{11}\
M_{\sun}$.  The color measured within a radius $r$ is plotted as a
function of that radius.  The polygon for a given snapshot illustrates
the range in colors as we view the galaxy from different angles.  The
size of the apertures used for the observations is indicated for
reference.

A color gradient is clearly present in all three phases, with the
color getting progressively bluer as we increase the aperture size.
The gradient is most pronounced during the merger event, and more so
for $\rfVJ$ than for $\rfUV$.  During the nuclear starburst ({\it
green polygon}), the difference (for a $1\farcs 5$ diameter aperture)
amounts up to 0.15 mag in $\rfUV$.  Given that our $\rfVJ$ estimates
were effectively derived from 3'' diameter apertures, a more
sophisticated simulation of the observing procedure could redden the
$\rfVJ$ colors by up to 0.1 mag.  For a given aperture size, the color
difference between the aperture and integrated photometry also
increases as we consider simulations of larger mass.  E.g., for a
merger that is five times more massive than that shown in Fig.\
\ref{colgrad.fig}, we find $(\rfUV$$)_{d=1\farcs 5}$ - $(\rfUV$$)_{\rm
total}$ reaches up to 0.25 mag and $(\rfVJ$$)_{d=3''}$ - $(\rfVJ$$)_{\rm
total}$ up to 0.6 mag.

We note that, since color apertures varied from object to object in
the observations, incorporating the details of the observing procedure
in our comparison is not straightforward.  Moreover, it would require
a proper treatment of smearing by the PSF, which was not applied in
Fig.\ \ref{colgrad.fig}.  Nonetheless, our analysis indicates that
color profiles are present in various degrees during different phases
of the merger scenario, and can contribute to the photometric
differences between simulated and real galaxies.  A study of color
profiles for high-redshift galaxies of different types, such as will be
enabled by the high resolution imaging with WFC3 onboard HST, will
provide additional constraints on the role of mergers in galaxy
evolution.

\subsubsection{Accounting for biases in SED modeling}
\label{biasSED.sec}
In comparing abundances of modeled and observed galaxies, it is
critical to apply identical selection criteria to both samples.  In
this paper, we work with mass-limited samples, where the stellar mass
of modeled galaxies is known from the simulation output, and that of
the observed galaxies is derived by fitting stellar population
synthesis templates to their multi-wavelength SED.  Wuyts et
al. (2009) tested the performance of the latter method by computing
mock high-redshift observations of the same simulations as used in
this paper, feeding them to the same SED modeling procedure as
described in \S\ref{masses.sec}, and comparing the derived masses with
the true values known from the simulation output.  The mass estimates
of merger remnants are very robust.  However, during phases of
merger-induced star formation systematic mass underestimates of 0.1 -
0.2 dex occur, with a tail toward more severe underestimates.  We
repeated our comparison of galaxy abundances, not using the true mass
of the modeled galaxies but the value derived from fitting their
virtually observed SEDs.  As expected from the results by Wuyts et
al. (2009), the modeled abundance of massive quiescent galaxies
remains nearly unaffected.  A significant fraction of simulated
star-forming galaxies drops out of the mass-limited sample, reducing
the modeled number density by a factor 1.6.  The mass density
decreases by a factor 1.8.  The shifts in number and mass density when
including the effects of mass underestimates of star-forming galaxies
in SED modeling are indicated with arrows in Fig.\ \ref{nrho.fig}.  We
conclude that, accounting for biases in SED modeling, the merger model
predicts that merger-induced star-forming galaxies above $\log M >
10.6$ can account for a third (quarter) of the total observed massive
star-forming population at $z
\sim 1.9$ (2.6).  Similar merger fractions among high-redshift
star-forming galaxies are found by kinematic studies of the SINS
survey (Shapiro et al. 2008).

\subsection {Dependence on stellar population synthesis and IMF}
\label{stelpop.sec}

As pointed out in \S\ref{VJ.sec}, the predicted rest-frame NIR
luminosities for a single stellar population of a given mass are
brighter for the M05 than for the BC03 stellar population synthesis
code.  Consequently, the mass estimates for observed galaxies with
ages between 0.2 and 2 Gyr are lower by about a factor 1.5 when
modeled with M05 instead of BC03 templates.  We indicated the
resulting systematic uncertainties in the observed and modeled number
and mass densities.  We find number densities for all samples
discussed in this paper to be typically two thirds and mass densities
to be $\sim$60\% of the value obtained with BC03 masses.  Whereas
using M05-based masses brings the observed abundance of star-forming
galaxies in agreement with the model prediction, it increases the
discrepancy between model and observations for the quiescent
population.  Furthermore, deviations from a Kroupa (2001) IMF would
change our results on number and mass densities of massive galaxies.
Recently, van Dokkum (2008) and Dav\'{e} (2008) presented evidence for
an evolving IMF that is more weighted to high mass stars at higher
redshift.  For the mass limit of $\log M > 10.6$ considered in this
paper, the general trend of such an evolving IMF would be to lower the
abundance of high-redshift galaxies above a certain mass limit, but is
not trivial to implement since the change in mass-to-light ratio would
depend on the galaxy's age.  Marchesini et al. (2008) are the first to
investigate the effect of bottom-light IMFs on the stellar mass
function, and find that it does not merely result in a shift.  The
precise shape of the stellar mass function depends on the
characteristic mass of the bottom-light IMF, and can in some cases,
perhaps counterintuitively, lead to abundances at the very high-mass
end ($\log M > 11.5$) that exceed those derived with a standard IMF.

Concerning the discrepancy in colors, this could be due to an
incorrect modeling of the stellar populations, rather than invalid
assumptions at the basis of the model (i.e., the one-to-one
correspondence between quasars and gas-rich mergers).  Apart from the
choice of stellar population synthesis code (see \S\ref{VJ.sec}), the
synthetic photometry depends on the attenuation law applied to each of
the stellar particles.  We note however that the use of a Milky
Way-like attenuation law from Pei (1992) leads to colors intermediate
between those based on the Calzetti et al. (2000) and SMC-like (Pei
1992) attenuation laws presented in this paper.  An attenuation law
that is less gray than that of the SMC would be required to reproduce
redder colors for dusty starburst galaxies.

Another stellar population parameter influencing the synthetic colors
is the metallicity of the gas and the stars.  In this paper, we
adopted initial gas metallicities derived from the closed box model
(Talbot \& Arnett 1971) for the 80\% gas fraction ($f_{\rm gas}$) at the
start of the simulation:
\begin {eqnarray}
Z_{\rm init} = -y \ln (f_{\rm gas})
\end {eqnarray}
where $y=0.02$ is the yield.  The simulation keeps track of the
subsequent evolution in the gas metallicity, and stellar metallicities
are based on the metallicity of the gas out of which they form.  If
the gas was pre-enriched, this would boost the optical depths and
redden the colors.  Evidence of high ($\sim Z_{\sun}$) metallicities
of massive high-redshift galaxies with red colors is given by van
Dokkum et al. (2004).  Repeating the post-processing of simulation
snapshots with $1 Z_{\sun}$ added to the gas and stellar
metallicities, we obtain colors that are 0.1 to 0.4 mag redder in
$U-V$ and 0.1 to 0.9 mag redder in $V-J$.  We note however that in
$V-J$ the largest increase occurs for blue galaxies and the color
distribution based on BC03 does not reach beyond $V-J \sim 1.8$.

\subsection {Radiative transfer}
\label {rt.sec}

The fact that we find a difference between the observed and modeled
$V-J$ colors of star-forming galaxies, does not necessarily mean that
the merger scenario (mergers triggering starbursts and quasar
activity, and leaving a quiescent remnant) should be abandoned.
Neither does it have to imply a failure of the hydro-simulations to
realistically model such a merger event.  Instead, and in fact more
plausibly, it might reflect the difficulty of translating physical
parameters stored in the GADGET-2 output into observables such as
colors and fluxes.  
So far, we followed Hopkins et al. (2005a) and Wuyts et al. (2009) to
compute the synthetic photometry assuming the cold gas clouds have a
negligible volume filling factor and ignoring the effect of
scattering.  In this subsection, we explore how the synthetic
photometry changes when adopting the recently developed radiative
transfer code SUNRISE (Jonsson 2006).  SUNRISE is a polychromatic
Monte Carlo code that uses a 3-D adaptive grid to treat arbitrary
geometries of emitting and absorbing/scattering media.  It
self-consistently treats dust re-emission and self-absorption by
iterating until the dust temperature converges.  This is particularly
important in the highly optically-thick central regions of merging
galaxies, and allows for future comparisons with longer wavelength
observations by MIPS/Spitzer, SCUBA (see, e.g., Narayanan et
al. 2009), and the upcoming Herschel mission.  The latest version
(Jonsson, P., Groves, B. \& Cox, T. J. in prep.)  includes a sophisticated
photoionization code MAPPINGS and subresolution model to account for
attenuation in the HII and photodissociation regions around young
stars (Groves et al. 2008).  Outside these birth clouds, SUNRISE
follows the light traveling through the hot phase of the ISM, assuming
a negligible volume filling factor of the cold phase clouds, as was
done in our line-of-sight attenuation code.  In addition to a
different treatment of the radiative transfer (e.g., the inclusion of
scattering), SUNRISE + MAPPINGS also make use of a different stellar
population synthesis code (Starburst99 by Leitherer et al. 1999).

\begin {figure*} [t]
\centering
\plotone{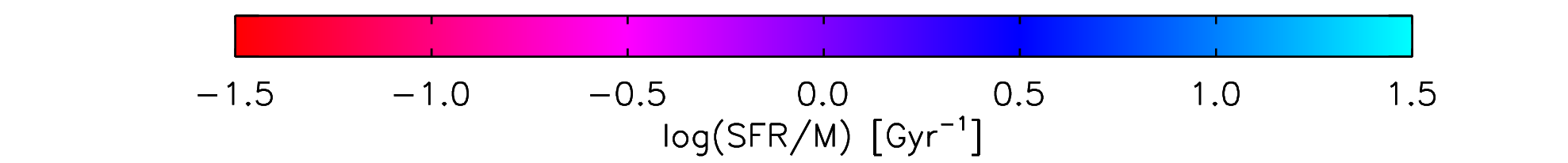} 
\plotone{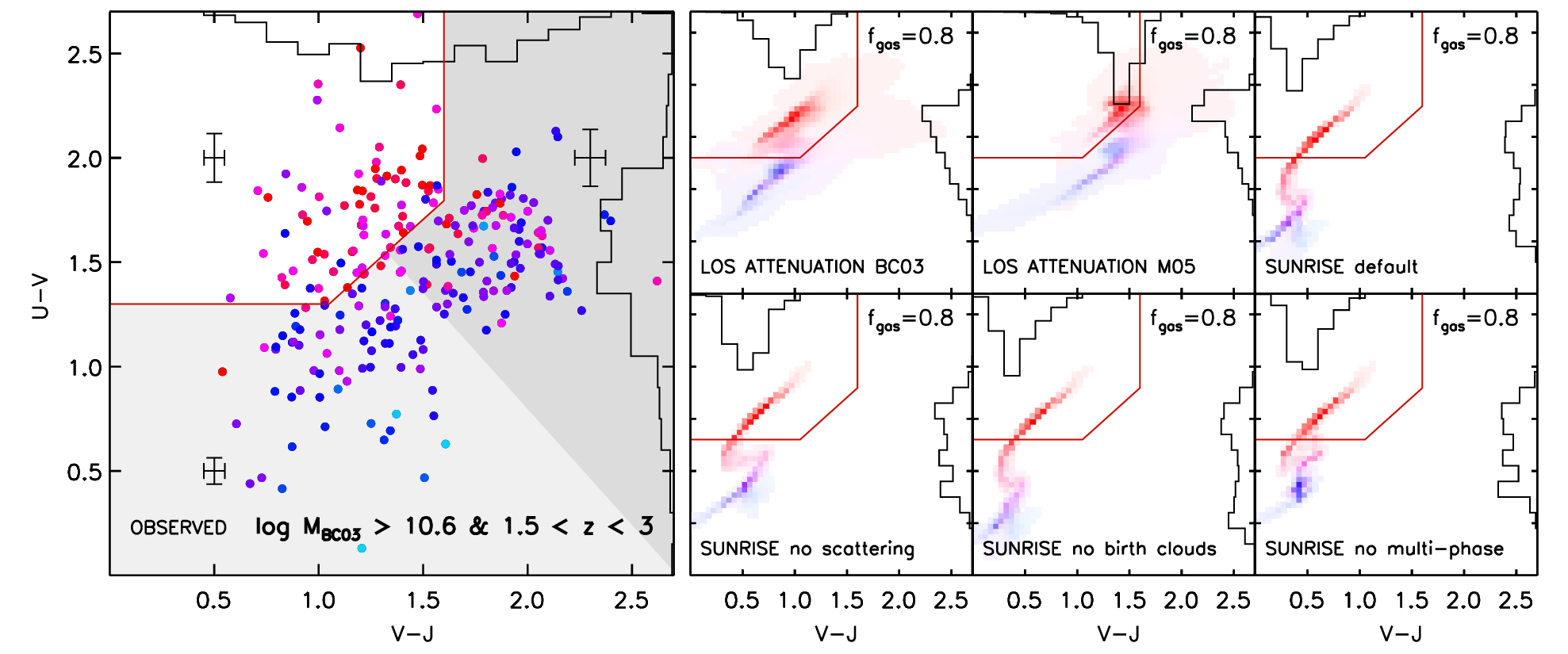} 
\plotone{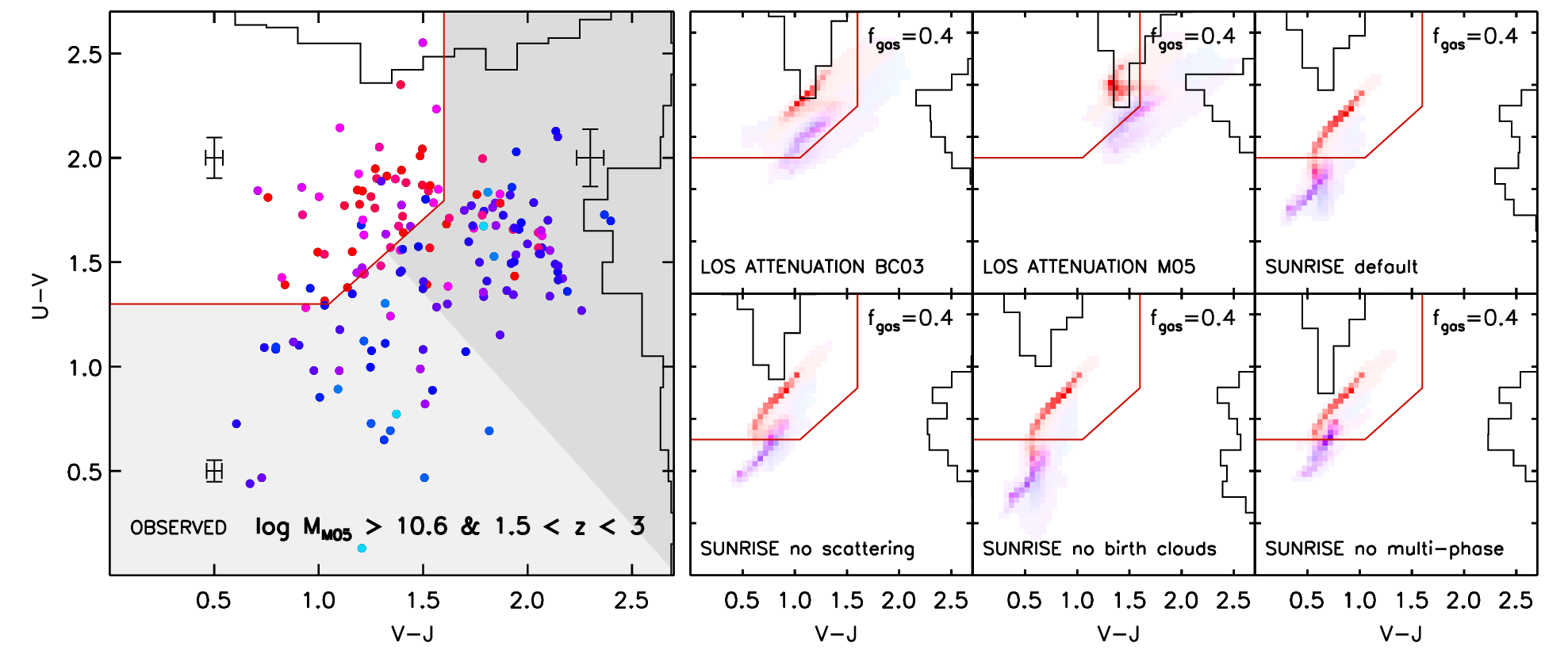} 
\caption{Rest-frame UVJ color-color diagram of massive galaxies at $1.5<z<3$ as
observed (big panels), and modeled with a variety of stellar
population synthesis codes and radiative transfer methods (small
panels).  Median error bars are plotted for observed galaxies enclosed
by the white, light-gray and dark-gray polygons.  Massive galaxies
exhibit a large range of specific star formation rates, both in the
observations and simulations.  The model colors depend on input
stellar population synthesis, method of radiative transfer, and
initial conditions of the simulations (e.g., initial gas fraction).
The model colors of quiescent galaxies are generally consistent with
those of their observed counterparts.  However, in none of the
realizations of the model photometry, the locus of star-forming
galaxies extends as far redward (in $\rfUV$ and particularly $\rfVJ$)
as the observations.
\label{SUNRISE.fig}}
\end {figure*}
We ran SUNRISE on the simulations used in this paper, using a SMC dust
model and identical initial conditions as adopted for our
line-of-sight attenuation calculations.  We adopted a birth cloud
covering fraction of 0.3, but note that increasing the covering
fraction to unity does not significantly alter our results.
Populating our model universe using the merger rate function of
\S\ref{methodology.sec}, we obtain the model color distribution for
the redshift interval $1.5<z<3$ shown in Fig.\ \ref{SUNRISE.fig}.  The
color distribution is plotted separately for the runs with gas
fraction $f_{\rm gas}=0.8$ and $f_{\rm gas}=0.4$ (see also
\S\ref{minor.sec}), and includes both the pre- and the post-quasar
phase.  For reference, we also include panels illustrating the color
distribution of massive observed galaxies in the same redshift
interval (with masses derived from BC03 and M05 models), and
realizations of the model photometry using our line-of-sight
attenuation code in combination with BC03 and M05 stellar population
synthesis templates.  In each of the panels, the color-coding
indicates the specific star formation rate.  Typical error bars in the
UVJ diagram of observed galaxies are drawn for the quiescent, and blue
and red star-forming galaxies separately (their respective regions in
UVJ space are outlined by the polygons).  Apart from the default
SUNRISE photometry, we also present realizations in which one of the
aspects of the radiative transfer is switched off.  This allows us to
disentangle the impact of scattering, birth clouds, and the
multi-phase breakdown of the ISM.

The default SUNRISE colors lie blueward of our line-of-sight
attenuation photometry of the same set of simulations, hence
articulating the difficulty of reproducing the colors of observed
dusty starbursts (objects in the dark-gray polygon of the observed
panels in Fig.\ \ref{SUNRISE.fig}).  Whereas -for the colors presented
in this diagram- the birth cloud model by Groves et al. (2008) has a
negligible impact, we note that accounting for scattering causes a
significant part of the blueward shift.  Making the assumption that
all gas is in the hot phase (i.e., switching off the multi-phase
breakdown) results in colors that are slightly redder during the
star-forming phases than the default SUNRISE photometry, but still
significantly bluer than the observed dusty starbursts.

Our analysis highlights the need to understand the details of
translating physical quantities to observables (i.e., stellar
population synthesis and radiative transfer) in order to effectively
constrain models.  Moreover, we note that -apart from the freedom in
initial gas fraction $f_{\rm gas}$, which for our isolated mergers is
set by hand- this issue applies equally, or perhaps more severely, to
cosmological simulations, where the spatial resolution is often too
limited to apply full radiative transfer.

\subsection {Dust distribution}
\label{dustdistribution.sec}
Possibly, the distribution of dust in the simulated galaxies might not
reflect reality.  A more efficient reddening would be obtained if a
foreground screen of obscuring material were present.  One possible
mechanism that could produce such a configuration on a galactic scale
is a large-scale wind.  The GADGET-2 code (Springel 2005b) used to run
the simulations in principle allows for such a phenomenon, but an
investigation of the velocity field of the gas in the simulations is
required to check whether such a wind is effectively taking place.

On much smaller scales, below the resolution of the simulations, a
more effective reddening might be expected from taking into account
that the molecular clouds in which new stars are formed, have a finite
lifetime.  So far, we ignored the role of birth clouds (except for the
comparison with SUNRISE radiative transfer, see \S\ref{rt.sec}),
essentially assuming that they are instantaneously dispersed when new
stars form.  Modeling the attenuation by HII and HI regions around
young stars and by the ambient ISM, Charlot \& Fall (2000) came up
with a simple recipe where the attenuation of light from young stars
is increased threefold with respect to the attenuation of the light
from old stars, until the birth clouds disperse after $\sim 10^7$ yr.
However, applying this recipe as a sub-grid model, the integrated SEDs
of the simulated galaxies do not always redden over the entire
optical-to-NIR wavelength range.

We find that, until the star formation rate drops shortly after the
final coalescence, a threefold extinction toward young stars ($<$ 10
Myr) reddens the integrated $\rfUV$ color by 0 - 0.15 mag.  During the
early merger phases, we observe a similar trend for the integrated
$\rfVJ$ color, with birth clouds causing a reddening of 0 - 0.1 mag.
However, during the nuclear starburst, the trend is reversed and
integrated $\rfVJ$ colors bluen by 0 - 0.15 mag.  The reason is that,
although young stars are intrinsically bluer, the column densities to
the nuclear region where the starburst takes place are so much higher
that, even ignoring the attenuation by birth clouds, the young
component to the integrated light is redder in $\rfVJ$ than the old
component.  The balance between reddening and dimming by birth clouds
subsequently reddens the already red young component, but downweights
its contribution to the integrated light, thus producing a bluer
overall color.  At later times, the contribution of young stars is
negligible, and so are the changes in the integrated colors when
applying the simple recipe for birth clouds.  

An alternative implementation, in which birth clouds around young
stars ($<$ 10 Myr) have a fixed $A_V = 3$ instead of additional
attenuation proportional to the line-of-sight column density through
the entire galaxy, suffers from the same down-weighting of the young
component to the integrated light.  Simple birth cloud models of the
kind described in this section seem insufficient to produce the colors
of dusty starbursts.  As described in \S\ref{rt.sec}, the more
sophisticated birth cloud model adopted by SUNRISE causes an effective
reddening of the $\rfUV$ and $\rfVJ$ colors, but is also insufficient
to account for the colors of observed dusty starbursts.

It is very well possible that the detailed dust distribution, on
scales far below the resolution of our simulations, plays a crucial
role in determining the galaxy colors.  The processes that govern the
distribution of dust are highly complex, involving wind escape from
massive stars, dust grain formation, and diffusion in the interstellar
medium.  Unfortunately, we are not able to address those issues with
self-consistent numerical modeling in our galactic-scale simulations,
at the moment.  Although the discussion in \S\ref{rt.sec} and this
subsection does not do justice to the full complexity of the problem,
it serves to highlight the uncertainties in the simulated colors.

\subsection {Merger parameters}
\label {minor.sec}
In addition, the discrepancy in colors might imply that the
simulations are not characteristic for the merger activity occuring in
the real universe.  E.g., from the shape of the stellar mass function
it can be expected that galaxies at the high-mass end are more likely
to merge with galaxies of lower rather than comparable mass (Khochfar
\& Silk 2006; Peng 2007).  Hopkins et al. (2006b) confirmed the
robustness of the model for quasar lifetimes and the derived merger
rate function against changes in various parameters of the merging
galaxies, such as gas fraction, orbital parameters and changes in the
mass ratio of the progenitors (considering 1:1, 2:1, 3:1, and 5:1 mass
ratios).  The conversion from a quasar birth rate to a spheroid birth
rate relies on a proper knowledge of the black hole - bulge mass
relation and how it evolves with redshift.  This source of uncertainty
will be reduced as more stringent observational constraints of the
scaling relation at high redshift become available.  Considering
progenitor mass ratios, Dasyra et al. (2006) find for a population of
local ULIRGs that still have 2 distinct nuclei that the typical mass
fraction is 1.5:1, close to equal-mass mergers.  In order to refine
the model predictions, a detailed study of minor merger simulations is
required to determine the minimum mass ratio required to trigger a
(low-luminosity) quasar phase.

Finally, all of the simulations used in this work are equal-mass
gas-rich mergers ($f_{\rm gas}=0.8$ at the start of the simulation).
In order to study the dependence of our results on the adopted initial
gas fraction $f_{\rm gas}$, we repeated our analysis with $f_{\rm gas}
= 0.4$ instead of $f_{\rm gas} = 0.8$.  Although further gas infall is
still not modeled, this test gives a crude but illustrative
characterization of the dependence on the available gas reservoir and
may represent differences induced by the gas accretion or evolutionary
history.  The $f_{\rm gas}=0.8$ runs represent a scenario in which
most of the stars are formed during the merger.  In the $f_{\rm
gas}=0.4$ runs on the other hand, the progenitor disks are more mature
and a relatively smaller fraction of the final stellar mass is formed
in episodes of merger-triggered star formation.  The star formation
history prior to the start of the simulation is assumed to be
constant, and the initial metallicities of gas and stars are set
according to a closed box model (see Wuyts et al. 2009 for more
details).  As a result, the $f_{\rm gas}=0.4$ runs do not only start
with a larger fraction of the baryonic mass in stars, but also with
more metal-rich gas ($0.9 Z_{\sun}$ at the start of the simulation)
and an initial stellar population having an older mass-weighted age
(on the order of 1 Gyr) and larger mass-weighted metallicity ($\sim
0.4 Z_{\sun}$).

In broad terms, our conclusions remain unaltered when lowering the
adopted $f_{\rm gas}$ from 0.8 to 0.4.  Especially the results
regarding the post-quasar phase seem robust, since by that time the
stellar mass evolution of the low and high gas fraction runs have
converged and most of the sensitivity to the initial conditions has
been washed out.  In other words, the post-quasar systems have
distinctly lower $SFR/M$ values than the earlier evolutionary stages,
their UVJ colors lie in the selection wedge that was designed to
select the observed quiescent galaxies, and, as for the $f_{\rm gas} =
0.8$ results, we find that the model prediction for their number and
mass density as derived from the observed quasar luminosity function
exceeds the observed abundance by a factor of $\sim 2.5$.

For the model predictions of the star-forming population, the
following dependencies on $f_{\rm gas}$ are observed.  First, in the
$f_{\rm gas}=0.4$ runs more stars already formed prior to the
merger-triggered star formation phase.  Some systems in the early
stages of merging, that in the $f_{\rm gas}=0.8$ realization had not
built up a sufficient amount of mass to enter the sample, will now
fall above the mass limit, hence boosting the model prediction and
reducing the discrepancy with respect to the abundance of observed
massive star-forming galaxies by a factor $\sim 1.7$.  Second, the
initial conditions affect the intrinsic photometry of simulated
merging galaxies by the increased age and metallicity of the stellar
population, and the attenuated photometry is affected also by the
smaller amount and larger metal-content of the gas.  The presence of
an old underlying population and relatively smaller build-up of new
stars during the merger explains why the tail towards blue $V-J$ and
especially blue $U-V$ colors (see Fig.\ \ref{UVhist.fig} and Fig.\
\ref{VJhist.fig}) is lacking in the color distribution realized with
$f_{\rm gas} = 0.4$.  For our default BC03 models, we find the $U-V$
color distribution for the entire (pre- and post-quasar) population to
range from 1 to 2 with the central 68\% interval bracketed by $1.3 <
U-V < 1.8$.  The $V-J$ distribution ranges from 0.5 to 1.6, with a
narrow central 68\% interval of $0.9 < V-J < 1.1$.  The dependence on
population synthesis code and attenuation law is similar as for the
higher $f_{\rm gas}$ runs.  E.g., using M05 models we find a $V-J$
distribution ranging from 1 to 2 (central 68\% interval $1.3 < V-J <
1.4$).  We conclude that the discrepancy with respect to the
optical-to-NIR colors of observed dusty starbursts is not alleviated
by adopting a different initial gas fraction.  This is also
illustrated in Fig.\ \ref{SUNRISE.fig}.

\subsection {Evolutionary history}
\label {accretion.sec}
Alternatively, it is possible that dusty starburst galaxies are not
triggered by mergers, but had a different evolutionary history.  Daddi
et al. (2007a) make this claim based on the long star formation
timescales of ULIRGs at high redshift, and the relatively tight
relation between SFR and stellar mass.  We note, however, that our
simulations of isolated disk galaxies with initial conditions
identical to those of the merger progenitors also fail to produce
actively star-forming systems with colors similar to dusty red
starbursts.  While CO interferometry by Tacconi et al. (2006, 2008)
shows key evidence that major merging is taking place in essentially
all SMGs, the merger fraction among the general $z \sim 2$
star-forming population may be lower.  Spatially-resolved kinematic
studies by the SINS survey (F\"{o}rster Schreiber et al. 2006b; Genzel
et al. 2006, 2008) and van Starkenburg et al. (2008) show a
significant number of rotating disks with high ($\sim 100 M_{\sun}
{\rm yr}^{-1}$) SFR but no evidence for recent or on-going major
merging.  Applying the method of kinemetry to the SINS sample, Shapiro
et al. (2008) finds a third of $z \sim 2$ star-forming galaxies to be
undergoing major merging, consistent with our model prediction.

From a theoretical perspective, recent cosmological hydrodynamic
simulations have suggested that most of the gas accretion at the
high-mass end takes place in cold flows along dark matter filaments,
with the streams being $\sim 50$\% smoothly flowing material, and the
other $\sim 50$\% in clumps of mass ratio $< 10:1$ with respect to the
accreting system (Keres et al. 2005; Dekel \& Birnboim 2006; Ocvirk et
al. 2008; Dekel et al. 2009).  Unlike major mergers, such flows could
keep the rotating disk configuration observed in many star-forming $z
\sim 2$ galaxies intact.  Using a semi-analytic model with star
formation and feedback recipes based on hydrodynamic simulations,
Somerville et al. (2008) also find that most of the global star
formation occurs in a quiescent mode, rather than in merger-induced
starbursts.  This could explain why our model prediction for the
abundance of merger-triggered star-forming galaxies accounts for only
a third of all observed star-forming galaxies at the high-mass end
(see \S\ref{SF.sec} and \S\ref{biasSED.sec}).

We conclude that, while our results are consistent with a scenario
where all massive quiescent galaxies at $1.5<z<3$ have formed by a
merging event that triggered quasar activity, it leaves room for other
mechanisms than major mergers contributing significantly to the $z
\sim 2$ star-forming population.  It would be interesting to 
scrutinize such alternative scenarios in the same way as presented
here.

\subsection {Mass loss and intergalactic environment}
\label {environment.sec}
Gas replenishment from mass loss and infall of gas from the
intergalactic environment could change the optical depths and thus the
reddening factors.  The simulations only take into account a small
amount of mass loss: 10\% of the gas mass converted into stars is
instantaneously returned to the interstellar medium, accounting for
short-lived stars that die as supernovae (Springel \& Hernquist 2003).
The total fraction of the mass lost by an aging single stellar
population with Salpeter (1955) IMF amounts to $\sim 30\%$ (BC03) and
is even higher ($\sim 50\%$) for more realistic IMFs such as Kroupa
(2001) or Chabrier (2003).  Furthermore, the simulations do not allow
for infall of primordial gas at later times.  Consequently, they
cannot prove that descendants of galaxies that once showed up in the
quasar luminosity function and after the shutdown of star formation
reached red colors, will remain quiescent forever.  Small amounts of
newly accreted gas triggering star formation may be enough to shift a
post-quasar galaxy outside the quiescent region of color-color space
defined by Eq.\ \ref{colsel.eq}, thus dropping their contribution to
the observed galaxy population of massive quiescent red galaxies.
Cosmological simulations at sufficient resolution might resolve this
problem.  At the very least, it would be interesting to test the
behavior of simulated merger remnants hosting a supermassive black
hole when a small but continuous gas supply is applied.

We note that a rejuvenation of part of the post-quasar population
would simultaneously improve the agreement with the observations for
both the star-forming and the quiescent galaxy abundance.

\subsection {Cosmic variance}
\label{variance.sec}
From the observational side, cosmic variance is the dominant source
random of uncertainties for the determination of the number and mass
density of massive galaxies.  At $z<2.5$, this is even the case when
including the wider area MUSYC survey (Marchesini et al. 2008).  This
is particularly true for the quiescent population, that shows a
stronger clustering than the star-forming one (Williams et al. 2009).
Surveys over a significantly larger area than FIRES and GOODS-South,
but probing similar depths at optical-to-MIR wavelengths, are required
to better constrain the fraction of massive quiescent galaxies that
post-quasar galaxies can account for.  The ultraVISTA survey by
Dunlop, Franx, Fynbo,\& LeF\`{e}vre will provide NIR imaging over half
of the 2 square degrees COSMOS field to an unprecedented depth.  In
combination with very deep IRAC imaging during the warm Spitzer
mission, such multi-wavelength surveys will allow to simultaneously
probe further down the mass function of quiescent galaxies, and reduce
the effect of cosmic variance.

\section {Summary}
\label{summary.sec}

We confronted the model by Hopkins et al. (2006b) with observations of
massive galaxies at $1.5<z<3$.  The model translates the observed
quasar luminosity function into the abundance of massive merging
galaxies and merger remnants.  We derived the synthetic photometry for
these systems from a set of binary merger SPH simulations by Robertson
et al. (2006a, 2006b) and T. J. Cox, with a range of masses, and
including stellar and AGN feedback.  We extracted mass-limited samples
of $1.5<z<3$ galaxies with $M > 4 \times 10^{10}\ M_{\sun}$ and $M >
10^{11}\ M_{\sun}$ from the FIRES+FIREWORKS and FIRES+FIREWORKS+MUSYC
surveys respectively.  We tested the model by comparing the predicted
number and mass densities, the $U-V$ and $V-J$ color distributions,
and $SFR/M$ versus mass distribution with our observations of massive
galaxies at $1.5<z<3$.

We find that the overall number density of galaxies with $M > 4 \times
10^{10}\ M_{\sun}$ in the FIRES and GOODS-South fields ($n
=4.0^{+1.3}_{-1.3} \times 10^{-4}\ Mpc^{-3}$ at $z \sim 1.9$ and $n =
3.3^{+1.1}_{-1.2} \times 10^{-4}\ Mpc^{-3}$ at $z \sim 2.6$) is
consistent within the uncertainties with the model prediction ($n =
4.8 - 5.5 \times 10^{-4}\ Mpc^{-3}$ at $z \sim 1.9$ and $n = 1.9 - 2.2
\times 10^{-4}\ Mpc^{-3}$ at $z \sim 2.6$).  Likewise, the results
obtained for the mass density are consistent: $\rho_* =
3.8^{+1.3}_{-1.2} \times 10^7\ M_{\sun}\ Mpc^{-3}$ at $z \sim 1.9$ and
$\rho_* = 2.8^{+0.9}_{-0.9}
\times 10^7\ M_{\sun}\ Mpc^{-3}$ at $z \sim 2.6$ for the observations
and $\rho_* = 5.4 - 6.0 \times 10^7\ M_{\sun}\ Mpc^{-3}$ at $z \sim
1.9$ and $\rho_* = 1.9 - 2.2 \times 10^7\ M_{\sun}\ Mpc^{-3}$ at $z
\sim 2.6$.

Separating massive galaxies by type, we find that the model photometry
of the post-quasar population coincides with the region of $U-V$
versus $V-J$ color-color space that was defined by Labb\'{e} et
al. (in preparation) to select quiescent red galaxies.  The modeled
number and mass densities of massive ($M > 4 \times 10^{10}\
M_{\sun}$) quiescent galaxies is consistent with the observations at
$z \sim 2.6$, but somewhat larger (2-3 times) than observed at $z \sim
1.9$.  The results based on the UVJ diagnostic diagram and on a
MIPS-based $SFR/M$ threshold are consistent.

We added the MUSYC survey to our sample, increasing the area by a
factor of 3.6, but by the shallower depth restricting our analysis to
$M > 10^{11}\ M_{\sun}$ galaxies.  From this sample, we derive
qualitatively similar results.

Although less constrained, the predicted abundances of galaxies with
merger-triggered star formation (according to their UVJ colors or
$SFR/M > 1/t_{\rm Hubble}$ can also account for a significant fraction
of the observed actively star-forming galaxies (one third when using
masses based on BC03 templates and taking into account biases in SED
modeling).  However, the predicted color distribution of star-forming
galaxies does not match the observations.  In particular the colors of
red ($V-J > 1.8$) dusty starburst galaxies are not reproduced.  We
suggest a number of explanations for the lack of dusty red starburst
galaxies in the model predictions.  Possible reasons are an incomplete
simulation of the observing procedure, differences in stellar
population properties or merger characteristics between the observed
and simulated galaxies, a different history for dusty starbursts than
a merger-triggered scenario, infall of additional gas (and dust) from
the intergalactic environment or mass loss, and a different
distribution of the dust, e.g., caused by the presence of large-scale
outflows or birthclouds around young stars.

Finally, we find hints of a pair excess at small angular scales,
further strengthening the hypothesis that mergers play a key role in
galaxy evolution.

We conclude that the star formation in remnants of merger simulations
is quenched abruptly, leading to colors that correspond well to those
of observed massive galaxies with a quiescent stellar population.
Using a merger rate derived from the observed quasar luminosity
function, we obtain number and mass densities of the quiescent
population at $1.5<z<3$ that are consistent within a factor of 2 to 3
with the observations.  Possibly, the overprediction at $z \sim 1.9$
suggests the need to include gas infall as refinement to the model.
The predicted abundance of merger-triggered star-forming systems
accounts for 30-50\% of the observed star-forming population, leaving
ample room for other star formation mechanisms than major merging.
The most serious challenge to the model is posed by the color
distribution of star-forming galaxies, which is not well reproduced.
The detailed dust distribution, on a galaxy-wide scale and/or scales
far below the resolution of the simulation, may well cause this
problem.  With this work, we hope to encourage further investigations
focussing on the translation of simulated physical quantities to
observables.

S. Wuyts and T. J. Cox gratefully acknowledge support from the
W. M. Keck Foundation.  B. E. Robertson gratefully acknowledges
support from a Spitzer Fellowship through a NASA grant administrated
by the Spitzer Science Center.

\begin{references}
{\small
\reference{} Adelberger, K. L., Steidel, C. C., Shapley, A. E., Hunt, M. P., Erb, D. K., Reddy, N. A.,\& Pettini, M. 2004, ApJ, 607, 226
\reference{} Bolzonella, M., Miralles, J.-M.,\& Pell\'{o}, R. 2000, A\&A, 363, 476
\reference{} Bower, R. G., Benson, A. J., Malbon, R., Helly, J. C., Frenk, C. S., Baugh, C. M., Cole, S.,\& Lacey, C. G. 2006, MNRAS, 370, 654
\reference{} Brammer, G. B., van Dokkum, P. G.,\& Coppi, P. 2008, ApJ, 686, 1503
\reference{} Calzetti, D., et al. 2000, ApJ, 533, 682
\reference{} Chabrier, G. 2003, ApJ, 586, L133
\reference{} Charlot, S.,\& Fall, S. M., ApJ, 539, 718
\reference{} Cimatti, A., et al. 2008, A\&A, 482, 21
\reference{} Cowie, L. L., Songaila, A., Hu, E. M.,\& Cohen, J. G. 1996, AJ, 112, 839
\reference{} Cox, T. J., Di Matteo, T., Hernquist, L., Hopkins, P. F., Robertson, B.,\& Springel, V. 2006a, ApJ, 643, 692
\reference{} Cox, T. J., Dutta, S. N., Di Matteo, T., Hernquist, L., Hopkins, P. F., Robertson, B.,\& Springel, V. 2006b, ApJ, 650, 791
\reference{} Croton, D. J. et al. 2006, MNRAS, 367, 864
\reference{} Daddi, E., Cimatti, A., Renzini, A., Fontana, A., Mignoli, M., Pozzetti, L., Tozzi, P.,\& Zamorani, G. 2004, ApJ, 617, 746
\reference{} Daddi, E., et al. 2007a, ApJ, 670, 156
\reference{} Daddi, E., et al. 2007b, ApJ, 670, 173
\reference{} Dale, D. A.,\& Helou, G. 2002, ApJ, 576, 159
\reference{} Dasyra, K. M., Tacconi, L. J., Davies, R. I., Genzel, R., Lutz, D., Naab, T., Burkert, A., Veilleux, S.,\& Sanders, D. B. 2006, ApJ, 638, 745
\reference{} Dav\'{e}, R. 2008, MNRAS, 385, 147
\reference{} Dekel, A.,\& Birnboim, Y. 2006, MNRAS, 368, 2
\reference{} Dekel, A., et al. 2009, Nature, 457, 451
\reference{} De Lucia, G.,\& Blaizot, J. 2007, MNRAS, 375, 2
\reference{} Di Matteo, T., Springel, V.,\& Hernquist, L. 2005, Nature, 433, 604
\reference{} Di Matteo, T., Colberg, J., Springel, V., Hernquist, L.,\& Sijacki, D. 2008, 676, 33
\reference{} Ferrarese, L.,\& Merritt, D. 2000, ApJ, 539, L9
\reference{} Fioc, M.,\& Rocca-Volmerange, B. 1997, A\&A, 326, 950
\reference{} F\"{o}rster Schreiber, N. M., et al. 2006a, AJ, 131, 1891
\reference{} F\"{o}rster Schreiber, N. M., et al. 2006b, ApJ, 645, 1062
\reference{} Franx, M., et al. 2000, The Messenger 99, pp. 20-22
\reference{} Franx, M., et al. 2003, ApJ, 587, L79
\reference{} Franx, M., van Dokkum, P. G., F\"{o}rster Schreiber, N. M., Wuyts, S., Labb\'{e}, I.,\& Toft, S. 2008, ApJ, 688, 770
\reference{} Gebhardt, K., et al. 2000, ApJ, 539, L13
\reference{} Genzel, R., et al. 2006, Nature, 442, 786
\reference{} Genzel, R., et al. 2008, ApJ, 687, 59
\reference{} Giacconi, R., et al. 2002, ApJS, 139, 369
\reference{} Giavalisco, M.,\& the GOODS Team 2004, ApJ, 600, L93
\reference{} Gingold, R. A.,\& Monaghan, J. J. 1977, MNRAS, 181, 375
\reference{} Granato G. L., De Zotti, G., Silva, L., Bressan, A.,\& Danese, L. 2004, ApJ, 600, 580
\reference{} Groves, B., Dopita, M. A., Sutherland, R. S., Kewley, L. J., Fischera, J., Leitherer, C., Brandl, B.,\& van Breugel, W. 2008, ApJS, 176, 438
\reference{} Hasinger, G., Miyaji, T.,\& Schmidt, M. 2005, A\&A, 441, 417
\reference{} Hopkins, P. F., Hernquist, L., Cox, T. J., Di Matteo, T., Martini, P., Robertson, B.,\& Springel, V. 2005, ApJ, 630, 705
\reference{} Hopkins, P. F., Hernquist, L., Cox, T. J., Di Matteo, T., Robertson, B.,\& Springel, V. 2006a, ApJS, 163, 1 
\reference{} Hopkins, P. F., Hernquist, L., Cox, T. J., Robertson, B.,\& Springel, V. 2006b, ApJS, 163, 50 
\reference{} Hopkins, P. F., Richards, G. T.,\& Hernquist, L. 2007, 654, 731
\reference{} Hopkins, P. F., Hernquist, L., Cox, T. J., Robertson, B.,\& Krause, E. 2007, ApJ, 669, 45 
\reference{} Hopkins, P. F., Hernquist, L., Cox, T. J., Dutta, S. N.,\& Rothberg, B. 2008, ApJ, 679, 156
\reference{} Hopkins, P. F., Cox, T. J., Dutta, S. N., Hernquist, L., Kormendy, J.,\& Lauer, T. R. 2009, ApJS, 181, 135
\reference{} Jonsson, P. 2006, MNRAS, 372, 2
\reference{} Jonsson, P., Cox, T. J., Primack, J. R.,\& Somerville, R. S. 2006, ApJ, 637, 255
\reference{} Kennicutt, R. C. 1998, ARAA, 36, 189
\reference{} Keres, D., Katz, N., Weinberg, D. H.,\& Dav\'{e}, R. 2005, MNRAS, 363, 2
\reference{} Khochfar, S.,\& Silk, J. 2006, ApJ, 648, L21
\reference{} Kriek, M., et al. 2006, ApJ, 649, 71
\reference{} Kriek, M., van der Wel, A., van Dokkum, P. G., Franx, M.,\& Illingworth, G. D. 2008, ApJ, 682, 896
\reference{} Kroupa, P. 2001, MNRAS, 322, 231
\reference{} Labb\'{e}, I., et al. 2003, AJ, 125, 1107
\reference{} Labb\'{e}, I., et al. 2005, ApJ, 624, L81
\reference{} Leitherer, C., et al. 1999, ApJS, 123, 3
\reference{} Li, Y., et al. 2007, ApJ, 665, 187L
\reference{} Lucy, L. B. 1977, AJ, 82, 1013
\reference{} Magorrian, J., et al. 1998, AJ, 115, 2285
\reference{} Maraston, C. 2005, MNRAS, 362, 799
\reference{} Maraston, C., Daddi, E., Renzini, A., Cimatti, A., Dickinson, M., Papovich, C., Pasquali, A.,\& Pirzkal, N. 2006, ApJ, 652, 85
\reference{} Marchesini, D., van Dokkum, P. G., Quadri, R., Rudnick, G., Franx, M., Lira, P., Wuyts, S., Gawiser, E., Christlein, D.,\& Toft, S. 2007, ApJ, 656, 42
\reference{} Marchesini, D., van Dokkum, P. G., F\"{o}rster Schreiber, N. M., Franx, M., Labb\'{e}, I.,\& Wuyts, S. 2008, ApJ, submitted, astro-ph/08111773v1
\reference{} Narayanan, D., Hayward, C. C., Cox, T. J., Hernquist, L., Jonsson, P., Younger, J. D.,\& Groves, B. 2009, submitted to MNRAS (astro-ph/09040004)
\reference{} Ocvirk, P., Pichon, C.,\& Teyssier, R. 2008, MNRAS, 390, 13260
\reference{} Papovich, C., et al. 2006, ApJ, 640, 92
\reference{} Pei, Y. C. 1992, ApJ, 395, 130
\reference{} Peng, C. Y. 2007, ApJ, 671, 1098
reference{} Quadri, R., et al. 2007, ApJ, 654, 138
\reference{} Quadri, R. F., Williams, R. J., Lee, K.-S., Franx, M., van Dokkum, P. G.,\& Brammer, G. B. 2008, ApJ, 685, L1
\reference{} Reddy, N. A., Erb, D. K., Steidel, C. C., Shapley, A. E., Adelberger, K. L.,\& Pettini, M. 2005, ApJ, 633, 748
\reference{} Reddy, N. A., et al. 2006, ApJ, 644, 792
\reference{} Richards, G. T., et al. 2005, MNRAS, 360, 839
\reference{} Robertson, B., Cox, T. J., Hernquist, L., Franx, M., Hopkins, P. F., Martini, P.,\& Springel, V. 2006a, ApJ, 641, 21
\reference{} Robertson, B., Hernquist, L., Cox, T. J., Di Matteo, T., Hopkins, P. F., Martini, P.,\& Springel, V. 2006b, ApJ, 641, 90
\reference{} Rocha, M., Jonsson, P., Primack, J. R.,\& Cox, T. J. 2008, MNRAS, 383, 1281
\reference{} Rudnick, G., et al. 2003, ApJ, 599, 847
\reference{} Sanders, D. B., Soifer, B. T., Elias, J. H., Madore, B. F., Matthews, K., Neugebauer, G.,\& Scoville, N. Z. 1988, ApJ, 325, 74
\reference{} Salpeter, E. E. 1955, ApJ, 121, 161
\reference{} Shapiro, K. L., et al. 2008, ApJ, 682, 231
\reference{} Somerville, R. S. 2004, in Multiwavelength mapping of galaxy formation and evolution, ed. R. Bender,\& A. Renzini (Berlin: Springer) (astro-ph/0401570)
\reference{} Somerville, R. S., Lee, K., Ferguson, H. C., Gardner, J. P., Moustakas, L. A.,\& Giavalisco, M. 2004, ApJ, 600, L171
\reference{} Somerville, R. S., Hopkins, P. F., Cox, T. J., Robertson, B. E.,\& Hernquist, L. 2008, MNRAS, 391, 481
\reference{} Springel, V.,\& Hernquist, L. 2003, MNRAS, 339, 289
\reference{} Springel, V., Di Matteo, T.,\& Hernquist, L. 2005a, ApJ, 620, L79
\reference{} Springel, V., Di Matteo, T.,\& Hernquist, L. 2005b, MNRAS, 361, 776
\reference{} Steidel, C. C., Adelberger, K. L., Shapley, A. E., Pettini, M., Dickinson, M.,\& Giavalisco, M. 2003, ApJ, 592, 728
\reference{} Tacconi, L. J., et al. 2006, ApJ, 640, 228
\reference{} Tacconi, L. J., et al. 2008, ApJ, 680, 246
\reference{} Talbot, R. J,\& Arnett, W. D., 1971, ApJ, 170, 409
\reference{} Taylor, E. N., et al. 2009, ApJS, in press (astro-ph/0903.3051)
\reference{} Ueda, Y., Akiyama, M., Ohta, K.,\& Miyaji, T. 2003, ApJ, 598, 886
\reference{} van Dokkum, P. G., et al. 2004, ApJ, 611, 703
\reference{} van Dokkum, P. G., et al. 2006, ApJ, 638, 59
\reference{} van Dokkum, P. G. 2008, ApJ, 674, 29
\reference{} van Starkenburg, L., van der Werf, P. P., Franx, M., Labb\'{e}, I., Rudnick, G.,\& Wuyts, S. 2008, A\&A, 488, 99
\reference{} Williams, R. J., Quadri, R. F., Franx, M., van Dokkum, P. G.,\& Labb\'{e} 2009, ApJ, 691, 1879
\reference{} Wuyts, S., et al. 2007, ApJ, 655, 51
\reference{} Wuyts, S., et al. 2008, ApJ, 682, 985
\reference{} Wuyts, S., Franx, M., Cox, T. J., Hernquist, L., Hopkins, P. F., Robertson, B. E.,\& van Dokkum, P. G. 2009, ApJ, 696, 348
\reference{} Yan, H. et al. 2004, ApJ, 616, 63
\reference{} Younger, J. D., Hayward, C. C., Narayanan, D., Cox, T. J., Hernquist, L.,\& Jonsson, P. 2009, MNRAS, 396, L66

}
\end {references}
















\end {document}